\documentclass{article}
\usepackage{a4wide}

\usepackage{latexsym,newdefns,amssymb,amsmath,stmaryrd,xspace}
\usepackage{rotating,url,sansmath,pdflscape,multirow,paralist,tikz}
\usepackage{setspace}
\usepackage{proof}

\newcommand{\analz}{\textsf{analz}\;}

\newcommand{\Agent}{\mathsf{Agent}}
\newcommand{\Nonce}{\mathsf{Nonce}}
\newcommand{\Stamps}{\mathsf{Timestamp}}
\newcommand{\Tags}{\mathsf{Tag}}

\newcommand{\msg}{\mathsf{msg}}
\newcommand{\sign}{\mathsf{sign}}

\newcommand{\Spy}{\mathsf{spy}}
\newcommand{\Server}{\mathsf{srv}}

\newcommand{\mencoding}{message-encoding}
\newcommand{\anaming}{agent-naming}
\newcommand{\sbinding}{session-binding}

\newcommand{\imp}{{\mathbb I}}
\newcommand{\icon}{{;^{\imp}}}

\newcommand{\fat}[1]{\left\{\hspace*{-0.03in}\left|{#1}\right|\hspace*{-0.03in}\right\}}
\newcommand{\ifat}[1]{\left\{\hspace*{-0.03in}\left|{#1}\right|\hspace*{-0.03in}\right\}^\imp}

\newcommand{\sqfat}[1]{[\!|{#1}|\;\,\!\!\!\!\!]}

\def\shrimp{\textsc{Shrimp}\xspace}

\def\numExp{{40}}
\def\cjlib{{20}}
\def\nummod{{11}}
\def\modbyhand{{4}}
\def\ktypeflaw{{5}}
\def\an{{N}}
\def\sbinA{{B}}
\def\sbinB{{B}}
\def\me{{E}}
\def\ok{{}}
\def\True{{T}}
\def\False{{F}}
\def\DontCare{{X}}
\def\numan{{12}}
\def\numme{{5}}

\def\numExito{{33}}
\def\depth{{\#}}
\def\Terms{{\mathsf{A}}}
\def\Keys{{\mathsf{K}}}   
\def\keys{{\mathit{K}}}   
\def\mess{{\mathit{M}}}   

\def\Atoms{{\mathsf{TK}}}

\def\Adapt{{\mathit{Ad}}}

\def\conc{;}

\newcommand{\Scope}{{\mathcal{EQ}}}
\newcommand{\Rms}{\mathsf{R}}

\newcommand{\sbtrm}{\sqsubseteq}
\newcommand{\Parts}{{\mathsf{Parts}}}
\newcommand{\Flat}{{\mathsf{fl}}}
\newcommand{\Analz}{{\mathsf{Analz}}}
\newcommand{\Synthz}{{\mathsf{Synthz}}}

\newcommand{\SI}[3]{{#1[#2, #3]}}

\newcommand{\nqed}{$\blacksquare$}

\newtheorem{definition}{Definition}
\newtheorem{lemma}[definition]{Lemma}

\newtheorem{example}{Example}

\begin{document}
\title{Strand-Based Approach to Patch Security Protocols}
\author{\hspace*{.7in}Dieter Hutter \hspace*{1.5in} Ra\'ul Monroy\\
  \small
  \begin{tabular}{ccc}
    German Research Center for Artificial Intelligence && Tecnol\'ogico de Monterrey\\ 
    Enrique Schmidt Str.~5 && Lago de Guadalupe Km.~3.5\\
    D-28359 Bremen, Germany && 52926, Atizap\'an, Mexico\\
    Tel.~+49(421)218 64 277 && Tel.~+52(55)5864 5316\\
    Fax.~+49(421)218 9864 277 && Fax.~+52(55)5864 5651\\
    \texttt{hutter@dfki.de} && \texttt{raulm@itesm.mx}
  \end{tabular}
  \normalsize
}

\date{}

\maketitle
\thispagestyle{empty}

\begin{abstract}
  In this paper, we introduce a mechanism that aims to speed up the
  development cycle of security protocols, by adding automated aid for
  diagnosis and repair.  Our mechanism relies on existing verification
  tools analyzing intermediate protocols and synthesizing potential
  attacks if the protocol is flawed. The analysis of these attacks
  (including type flaw attacks) pinpoints the source of the failure
  and controls the synthesis of appropriate patches to the protocol.
  Using strand spaces~\cite{strand-spaces}, we have developed general
  guidelines for protocol repair, and captured them into formal
  requirements on (sets of) protocol steps.  For each requirement,
  there is a collection of rules that transform a set of protocol
  steps violating the requirement into a set conforming it.  We have
  implemented our mechanism into a tool, called \shrimp. We have
  successfully tested \shrimp on numerous faulty protocols, all of
  which were successfully repaired, fully automatically.
\end{abstract}

\section{Introduction}
\label{sec:intro}

A security protocol is a protocol that aims to establish one or more
security goals, often a combination of authentication, confidentiality
or non-repudiation.  Security protocols are critical applications,
because they are crucial to provide key Internet services, such as
electronic-banking. So, security protocols are thoroughly studied to
guarantee that there does not exist an interleaving of protocol runs
violating a security goal, called an \emph{attack}.  Designing a
security protocol is, however, error-prone.  Although security
protocols may be extremely simple, consisting of only a few steps,
they are difficult to get right.  Formal methods have been
successfully used to identify subtle assumptions underlying a number
of faulty protocols, yielding novel attacks.

In this paper, we introduce a mechanism that aims to speed up the
development cycle of security protocols, by adding automated support
for diagnosis and repair.  Our mechanism has been especially designed
to fix protocols that are susceptible to an attack of the full class
replay~\cite{Syverson94}.\footnote{\onehalfspacing A \emph{replay
    attack} is one where a valid message is maliciously repeated in
  other (not necessarily different) session; so, the active
  participation of a penetrator is required (c.f.~the Dolev-Yao
  penetrator model~\cite{DY83}).} This attack class includes a number
of sub-classes; some are well known, such as reflection, and unknown
key share, but others have passed slightly ignored, like type
flaw.\footnote{\onehalfspacing A \emph{type flaw attack} is one where
  a participant confuses a (field of a) message containing data of one
  type with a message data of another.}

\paragraph{Overview of Approach for Comparison}

There exist several approaches to security protocol development,
ranging from the systematic generation of a protocol (called
\emph{protocol synthesis}) to the transformation of a given protocol
into one that is stronger, up to some specified properties (called
\emph{protocol compiling}).  Protocol synthesis (blindly) generates a
candidate protocol, and then tests it to see if it complies with the
intended security requirements. By contrast, protocol compiling
(blindly) transforms an input protocol, by wrapping it with explicit
security constructs; thus, protocol compiling barely depends on the
security guarantees provided by the input protocol, if any.  Our
method is in between, and follows the formal approach to software
development; it is applied after a failed verification attempt, in
order to spot a protocol flaw, and suggest a candidate patch.

To compare the relative value of our approach against that of all
these competitors, we suggest to use efficiency of a given output
security protocol. There are three common criteria for measuring this
dimension~\cite{KatzY07}:
\begin{compactdesc}
\item[\textbf{Round complexity},] the number of rounds until the
  protocol terminates. A round is a collection of messages that can be
  simultaneously sent by parties, assuming that the adversary delivers
  all these messages intact, immediately, and to the corresponding
  party.
\item[\textbf{Message complexity},] the maximum number of messages
  sent by any single party.
\item[\textbf{Communication complexity},] the maximum number of bits
  sent by any single party.
\end{compactdesc}
We shall have more to say about this later on in the text, when
comparing our method against rival techniques (see
Section~\ref{sec:related-work}.)

\paragraph{Contributions of Paper}

In this paper, we make four key contributions. First, we provide a
fully automated repair mechanism, which supersedes and surpasses our
previous methods (see~\cite{shrimp,type-flawHM}.)  The new mechanism
has been entirely developed within strand spaces~\cite{strand-spaces},
and it is no longer heuristic-based. We have used strand spaces to
provide a full characterization of a wide class of protocol attacks,
and our repair strategies, in a way amenable to mechanization. Our
mechanism for protocol repair has been successfully tested on a number
of faulty protocols collected from the literature. Second, we provide
a handful of basic principles for protocol repair, stemmed from our
insights on our strand spaces characterization of a failed
verification attempt. Our repair principles are simple and intuitive,
and follow from a straight formalization of an interleaving of partial
protocol runs, and the way messages are used to achieve an attack.
Third, we provide an extension of the theory of strand spaces, which
encapsulates issues about the implementation of messages, hence,
providing a neat border between reasoning about the representation of
a message, from reasoning about its actual form (a byte stream). Our
extension consists of an equivalence relation over messages, and a
distinctive notion of message origination. With it, we have been able
to capture type flaw attacks, using the constructions of the strand
spaces calculus. What is more, our abstraction of messages enables the
transparent application of our strategies for protocol repair; that
is, we can fix protocols that are subject to a replay attack because a
protocol message can be mixed up with other one, without considering
whether message confussion exploits a type flaw or not.  Fourth, we
provide a comparison of protocol repair against other prominent
approaches for protocol development, concluding that our approach
gives the designer a better insight for protocol development, guiding
the conception for the r\^ole of messages and their composition.

\paragraph{Overview of Paper}
The remaining of the paper is organized as follows: first, we overview
our approach to protocol repair (see Section~\ref{sec:approach}).
Then, we introduce notation and key, preliminary concepts (see
Section~\ref{sec:prelim}). Next, we introduce and discuss the
soundness of our extended version of strand spaces, suitable to
capture type flaw attacks (see Section~\ref{sec:itraces}). Then, after
describing the kinds of flaws we want to automatically repair (see
Section~\ref{sec:principles}), we provide a method for automatic
protocol repair (see Section~\ref{sec:rules}), and the results
obtained from an experimental test (see Section~\ref{sec:results}). We
conclude the paper, after comparing related work and giving directions
to further work (see Sections~\ref{sec:related-work}
and~\ref{sec:conclusions}).


\section{Protocol Repair: General Approach}
\label{sec:approach}

\begin{figure}[htb]
  \hrule
  \begin{center}
    \includegraphics[width=0.7\textwidth]{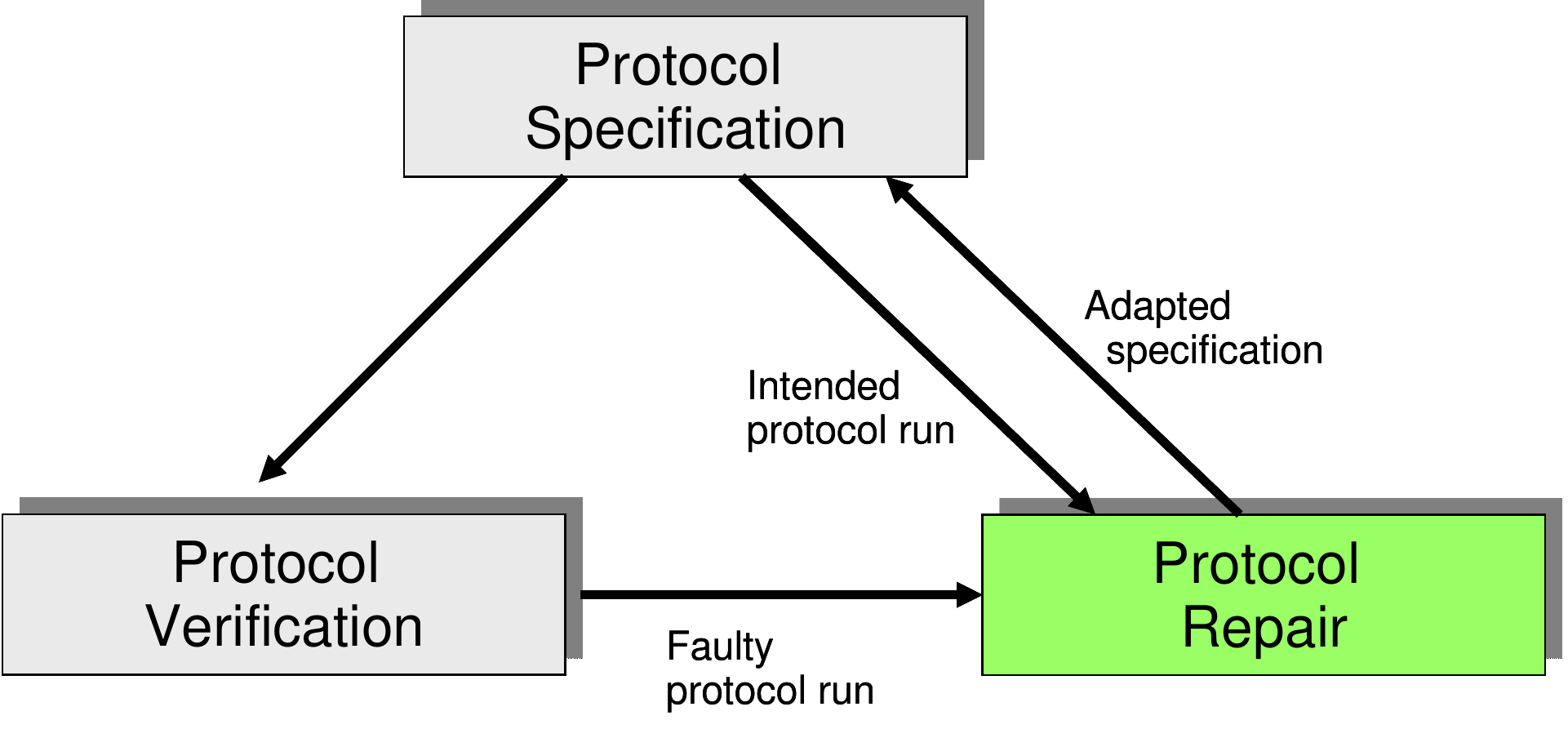}
  \end{center}
  \hrule
  \caption{Overview of the general approach}
  \label{Fig:Overview}
\end{figure}

Our approach to protocol repair involves the use of two tools (see
Fig.~\ref{Fig:Overview}). One tool is a protocol verifier
(e.g.~OFMC~\cite{ofmc,alg-spy-ded}, ProVerif~\cite{proverif}, or
Cl-Atse~\cite{Turuani06}), which is used to analyze the protocol at
hand and to yield an attack on that protocol, if the protocol is
faulty. The other tool, called \shrimp, embodies our mechanism for
protocol repair. \shrimp compares the attack against a run of the
protocol where the penetrator does not participate, called a
\emph{regular run}. This comparison usually spots differences that
indicate where the protocol went wrong.  Depending on these
differences, \shrimp offers various candidate patches, each of which
modifies the protocol in a specific way.  Some patches may rearrange
parts of an individual message or enrich it with additional
information, while others may alter the flow of messages in the
protocol.

Our approach to protocol repair is iterative. Since we proceed from a
local perspective to fix a problem that has become apparent during
attack inspection, we cannot guarantee that a fix will remove all
flaws from the protocol.  So, the mended protocol is sent back to the
verification tool for reanalysis.  If this tool still detects a bug,
either because this other bug was already present in the original
protocol (as often is) or because it was introduced by our method (as
has never happened so far), we iterate, applying our repair strategies
on the mended protocol.

Our patches change the protocol in a very restricted and conservative
way, since we have no explicit representation of the intended purpose
of a protocol.  For example, suppose that to prevent a replay attack
we have to change the structure of an individual message in a protocol
so that it uniquely determines the protocol run, as well as the
protocol step it originates from.  To preserve the semantics of that
message, we modify it only in such a way that the new message
preserves both encryption (it is encrypted in the same way) and
information (it contains all information present in the original
message), with respect to the knowledge of some keys.

Attack analysis and protocol repair are driven by a handful of
informal guidelines for protocol design. 
We have translated these principles into formal requirements on sets
of protocol steps.  For each requirement, there is a collection of
rules that transform a set of protocol steps violating the requirement
into a set conforming it.  The correction of security protocols
incorporates the use of several of these rules.  However, protocol
patches are not independent; so a rule requires preconditions to be
applicable and should guarantee postconditions once it has been
applied.


We have developed our mechanism within (our extension to) the theory
of strand spaces~\cite{strand-spaces}.  Thus, the specification of a
protocol and of one of its associated attack have both to be given
using the strand space notation. Extra machinery is required to
translate a given protocol specification so it is suitable for the
chosen verification tool (see Fig.~\ref{Fig:Overview}). 
Also, extra
machinery is required to translate the attack output by that
verification tool, if the protocol is faulty, so that the attack can
be used by \shrimp.  We will see later on in the text that since a
protocol specification and a protocol attack must each form a bundle,
this translation is straightforward.  However, care needs to be taken
when the attack to the protocol is type-flaw, since in this case the
translation needs to deal with the introduction of special
construction blocks that justify the type flaw attack. We call these
construction blocks implementation traces, as shall be seen in
Section~\ref{sec:itraces}. A preliminary version of \shrimp is
available at
\small\url{http://homepage.cem.itesm.mx/raulm/pub/shrimp/}\normalsize.

\section{Formal Preliminaries}
\label{sec:prelim}

We now recall standard concepts and notations about security
protocols, in general, and from strands
spaces~\cite{THG98,strand-spaces,auth-tests}, in particular.  For a
thorough introduction to the strand spaces, though, readers are
referred to~\cite{strand-spaces}.

As usual, we consider messages, $\Terms$, as the set of terms, which
is \emph{freely} generated from the sets of atomic messages, $\Atoms$,
consisting of nonces, $\Nonce$, timestamps, $\Stamps$, agent names,
$\Agent$, tags, $\Tags$, and keys, $\Keys$, using concatenation,
$m_1\conc m_2$, and encryption, $\fat{m}_k$, with $k\in\Keys$ and $m,
m_1,m_2\in\Terms$.

We assume two functions. One maps principals, $a,b,\ldots$, to their
public keys, $k_a,k_b,\ldots$, and the other a pair of principals,
$\Tuple{a}{b}$, to their symmetric, shared key, $k_{ab}$.  The set of
keys, $\Keys$, comes with an inverse operator, mapping either each
member of a key pair for an asymmetric cryptosystem to the other,
$(k_a)^{-1}=k^-_a$, or each symmetric key to itself,
$(k_{ab})^{-1}=k_{ab}$.

$\Parts_{\keys}$ maps a message $t\in\Terms$ to the set containing all
the subterms of $t$ that are accessible knowing the set of keys,
$\keys$; i.e.~it is the least set satisfying: $t\in\Parts_{\keys}(t)$,
$\Parts_{\keys}(t_i)\subseteq\Parts_{\keys}(t_1;t_2)$, for
$i\in\{1,2\}$, and
$\Parts_{\keys}(t)\subseteq\Parts_{\keys}(\fat{t}_k)$, if
$k^{-1}\in\keys$.  $\Parts(t) =
\Parts_{\Keys}(t)$ defines the set of all subterms.  We write $t
\sbtrm_{\keys} t'$ (and $t\sbtrm t'$, respectively) to denote $t\in
\Parts_{\keys}(t')$ (and $t\in\Parts(t')$, respectively). 

Let $t$ and $\keys$ be a message and a set of keys, respectively.
Further, let $\mess_0, \mess_1, \ldots$ be a sequence of sets of
messages and $\keys_0, \keys_1, \ldots$ be a sequence of sets of keys,
such that $\keys_0=\keys$, $\mess_0=\Parts_{\keys}(t)$, $\keys_{i+1} =
\keys_i\cup(\Keys \cap \mess_{i})$, and $\mess_{i+1} =
\bigcup_{t'\in\mess_i}\Parts_{\keys_{i+1}}(t')$.  Since a message
contains only finitely many keys, there is an $i_0$ with $\keys_j =
\keys_{j+1}$, for all $j \ge i_0$. Therefore, $\mess_0,\mess_1,\ldots$
has a fixed-point, $\mess_{\infty}$, which we call
$\Analz_{\keys}(t)$.  Let $M$ be a set of messages; then $\Synthz(M)$
is the smallest set such that: $M\subseteq\Synthz(M)$; if
$m_1,m_2\in\Synthz(M)$, then $m_1;m_2\in\Synthz(M)$; and if
$m\in\Synthz(M)$ and $k\in\Synthz(M)$, then $\fat{m}_k\in\Synthz(M)$.

Let $\Flat(t)$ be the set of all messages in a message $t$ obtained by
flattening the top-level concatenations; i.e.~$\Flat(t) = \{ t' \in
\Parts_{\emptyset}(t) | t' \in \Atoms \text{ or } \exists m, k.\; t' =
\fat{m}_k \}$. Whenever $m\in\Flat(t)$, $m$ is said to be a
\emph{component} of $t$. In an abuse of notation, we write $\Atoms(m)$
for the set of atomic messages in $m$; i.e.~$\Atoms(m) = \{t\in\Atoms
| t\in\Parts_{\Keys}(m)\}$. We also extend $\Atoms(\cdot)$ to a
homomorphism over sets of terms in the expected manner.


A strand represents a principal's local view on a protocol.  Thus, a
\emph{strand} $s$ is a sequence of directed messages (called
\emph{nodes}), $\pm t_1\Rightarrow_s\pm t_2,\ldots\Rightarrow_s\pm
t_n$, each of which contains the information whether the message is
received from the outside (indicated by the sign ``-''), or sent by
the principal (indicated by ``+'').  Accordingly, we call a node
either \emph{positive} or \emph{negative}.  The relation
$\Rightarrow_s$ connects consecutive messages in strand $s$.
$\Rightarrow_s^+$ and $\Rightarrow_s^*$ are respectively used to
denote the transitive, and the transitive-reflexive closure of
$\Rightarrow_s$. We write $\Rightarrow_{\mathcal{T}}$ for the union of
$\Rightarrow_s$, for all $s\in\mathcal{T}$. Notice, however, that,
when understood from the context, we shall simply write $\Rightarrow$
to refer to $\Rightarrow_s$.

Let $v=\Tuple{s}{i}$ denote the $i$-th node of a strand $s$. Then, we
respectively use $\msg(v)$ and $\sign(v)$ to denote the message, and
the event, sending or reception, associated with node $v$. We write
$\depth(s)$ to stand for the length of a strand $s$; thus, $1\leq
i\leq\depth(s)$, whenever $\Tuple{s}{i}$ denotes the $i$-th node of
strand $s$.  

The abilities of a Dolev-Yao penetrator are characterized by means of
a set of building blocks, called penetrator strands. There are eight
penetrator strands: \textbf{(K)ey}: $\langle +k\rangle$, provided that
$k\in\keys_\mathsf{P}$, with $\keys_\mathsf{P}$ being the set of keys
initially known to the penetrator; \textbf{(T)ee}:
$\Tuple{-m}{+m,+m}$; \textbf{(F)lush}: $\langle -m\rangle$;
\textbf{(M)essage}: $\langle +m\rangle$, provided that $m\in\Agent$ or
$m\in\Nonce$; \textbf{(C)oncatenation}: $\Tuple{-m_1}{-m_2,+m_1\conc
  m_2}$; \textbf{(S)eparation}: $\Tuple{-m_1\conc m_2}{+m_1,+m_2}$;
\textbf{(E)ncryption}: $\Tuple{-k,-m}{+\fat{m}_k}$; and finally
\textbf{(D)ecryption}: $\Tuple{-\fat{m}_k}{+k,+m}$.
\label{pag:penetrator-strands}

A \emph{strand space} $\Sigma$ denotes a set of strands.  Given two
strands, $s$ and $s'$, with $s\neq s'$,
$\Tuple{s}{i}\rightarrow\Tuple{s'}{j}$ represents inter-strand
communication from $s$ to $s'$. It requires the nodes messages to be
equal, i.e.\ $\msg(\Tuple{s}{i})=\msg(\Tuple{s'}{j})$ but having
complementary signs, i.e.\ $\sign(\Tuple{s}{i}) = +$ and
$\sign(\Tuple{s'}{j})=-$.

A \emph{bundle} is a composition of (possibly incomplete) strands and
penetrator traces, hooked together via inter-strand communication.
Formally, it is a finite, acyclic graph
$\mathcal{B}=\Tuple{V}{(\rightarrow\cup\Rightarrow)}$, such that for
every $v_2\in V$, the two following conditions hold: i) if
$\sign(v_2)=-$, then there is a unique $v_1\in V$ with $v_1\rightarrow
v_2$; and ii) if $v_1\Rightarrow v_2$ then $v_1\in V$ and
$v_1=\Tuple{s}{i}$ and $v_2=\Tuple{s}{i+1}$. $\prec_{\cal B}$ and
$\preceq_{{\cal B}}$ denote respectively the transitive and the
transitive-reflexive closure of $(\rightarrow\cup\Rightarrow)$.

Let $s_a$ be the strand of participant $a$. If $a$ is honest, then the
strand $s_a$, as well as each individual strand node,
$\Tuple{s_a}{i}$, is said to be \emph{regular}. Otherwise, it is said
to be \emph{penetrator}.  Let $v =
\Tuple{s_a}{j}$ be some node in the strand of participant $a$; then,
we use $\Agent(v) = a$ to denote the strand participant name.

The pair $(\Sigma,\mathcal{P})$ is said to be an \emph{infiltrated
  strand space}, whenever $\Sigma$ is a strand space and
$\mathcal{P}\subseteq\Sigma$, such that every $p\in\mathcal{P}$ is a
penetrator strand.  In an infiltrated strand space,
$(\Sigma,\mathcal{P})$, the penetrator can build masquerading messages
using \textbf{M}, \textbf{K}, \textbf{F}, \textbf{T}, \textbf{C},
\textbf{S}, \textbf{E}, and \textbf{D}, only. Notice that penetrator
traces of type \textbf{M} cannot suddenly output an unguessable nonce,
which are modelled using origination assumptions.

A bundle is said to be \emph{regular}, if it contains no penetrator
strands, and it is said to be \emph{penetrator}, otherwise. A message
$m$ is said to be a component of a node $v$ if it is a component of
$\msg(v)$.  A node $v$ is said to be a \textbf{M}, \textbf{K},
\ldots~node, if it lies on a penetrator strand with a trace of kind
\textbf{M}, \textbf{K}, \ldots. 

The following section is entirely devoted to the extension of strand
spaces that allows the capture of type-attack flaws.

\section{Extending Strand Spaces to Capture Type Flaw Attacks}
\label{sec:itraces}

The strand space approach, \cite{strand-spaces}, see above, introduces
messages as a freely generated datatype, built up on atomic messages,
like nonces, names, or keys, with the help of the two constructors,
namely: concatenation and encryption. Hence, messages are considered
to be different if their syntactical representations (as constructor
ground terms) are different. However, in practice, messages are
typically implemented as byte-streams, and their parsing as messages
may not be unique: there might be two different messages that share
the same implementation as a byte-stream. Thus, the expectation of the
receiver about a byte-stream controls the interpretation it gives to
the message. Put differently, a penetrator can fake a message (or
parts thereof) by (re-)using messages of the wrong type but with
suitable implementation.
\begin{example}
  \label{ex:WooLamPi}
  For example, consider the Woo Lam $\pi_1$ protocol~\cite{woolam},
  given as a bundle below (there, and henceforth, Init, Resp, and Serv
  stand for the initiator, responder, and the server, respectively):

  \begin{eqnarray*} 
    \begin{array}{ccccc}
      + a & \xrightarrow{\hspace*{0.1in}} & - a \\
      \Downarrow & & \Downarrow \\
      - n_b & \xleftarrow{\hspace*{0.1in}} & + n_b \\
      \Downarrow & & \Downarrow \\
      +\fat{a\conc b\conc n_b}_{k_{as}} & \xrightarrow{\hspace*{0.1in}} & - \fat{a\conc b\conc n_b}_{k_{as}} \\
      & & \Downarrow \\
      & & +\fat{a\conc b\conc\fat{a\conc b\conc n_b}_{k_{as}}}_{k_{bs}} &  \xrightarrow{\hspace*{0.1in}} 
      & - \fat{a\conc b\conc\fat{a\conc b\conc n_b}_{k_{as}}}_{k_{bs}} \\
      & & \Downarrow & & \Downarrow \\
      & & - \fat{a\conc b\conc n_b}_{k_{bs}} & \xleftarrow{\hspace*{0.1in}} & + \fat{a\conc b\conc n_b}_{k_{bs}} \\[2ex]
      Init && Resp && Serv  
    \end{array} 
  \end{eqnarray*} 
  This protocol is vulnerable to a type flaw attack, when the
  implementation of nonces can be confused with the implementation of
  encrypted messages, giving rise to the following attack:
  \begin{eqnarray*}
    \begin{array}{rrcll}
      1. & \Spy(a) & \rightarrow & b & : a\\
      2. & b & \rightarrow & \Spy(a) & : n_b\\
      3. & \Spy(a) & \rightarrow & b & : n_b\\
      4. & b & \rightarrow & \Spy(\Server) & : \fat{a\conc b\conc n_b}_{k_{bs}}\\
      5. & \Spy(\Server) & \rightarrow & b & : \fat{a\conc b\conc n_b}_{k_{bs}}
    \end{array}
  \end{eqnarray*}
  That $n_b$ and $\fat{a\conc b\conc n_b}_{k_{as}}$ can be confused is
  not unreasonable, since the receiver, $b$, does not know the key,
  $k_{as}$, shared between $a$ and $s$, and it is therefore not able
  to realize the hidden structure of the message under
  encryption.\hfill\nqed
\end{example}

Strand spaces cannot represent the attack above, since terms are
freely generated. Hence any guarantees we might obtain from this
theory could make use of the freeness axioms, and so could not be
transferred to a situation where they do not hold. So, in what
follows, we will extend the strand space notation to deal with type
flaw attacks.

\subsection{Theories of Message Implementation}

In a first step, we have to specify the implementation of messages,
which we use to represent a run of the protocol in practice. We assume
that there is some translation function $\imp$ that maps any message
$t$ (given in the freely generated datatype) to its implementation
$\imp(t)$.  Furthermore, we assume also that there are implementations
$\ifat{\_}$ and $\icon$ for concatenation and encryption on the
implementation level, such that $\imp$ is a homomorphism from messages
to their implementations. We also assume that $\imp$ denotes a
faithful representation on individual message types, i.e. we can
distinguish the implementations of, for instance, two agents or two
nonces (while still a nonce and a key can share a common
implementation). The following definition formalizes this idea and
provides the minimal requirements to an implementation.
\begin{definition}[Implementation, $\approx$]\label{def:impl}
  Let $\Rms$ be a set and let $\imp$ be a mapping from $\Terms$ to
  $\Rms$.  $(\imp, \Rms)$ is an \emph{implementation} of $\Terms$ iff
  the following holds.
  \begin{eqnarray}
    \forall x, y\in\Terms. \; \imp(x) \icon \imp(y) = \imp(x; y)\label{ax1}\\
    \forall x, y\in\Terms. \; \ifat{\imp(x)}_{\imp(y)} = \imp(\fat{x}_y)\label{ax2}\\
    \forall x, y\in\Agent. \;  \imp(x) = \imp(y) \to x = y\label{ax3}\\  
    \forall x, y\in\Keys. \;  \imp(x) = \imp(y) \to x = y\label{ax4}\\  
    \forall x, y\in\Nonce. \;  \imp(x) = \imp(y) \to x = y\label{ax5}\\
    \forall x, y\in\Stamps. \;  \imp(x) = \imp(y) \to x = y   \label{ax6}
  \end{eqnarray}
  where $x$ and $y$ are all meta-variables. We write
  $m_1\approx_{\imp} m_2$ iff $\imp(m_1)=\imp(m_2)$. If $\imp$ is
  fixed by the environment we simply write $m_1\approx m_2$, instead
  of $m_1\approx_{\imp} m_2$
\end{definition}		
		
For a typical setting, we might need to extend
Definition~\ref{def:impl} with additional axioms specifying in more
detail how messages and their operations are implemented.  For
example, one might like to formalize an implementation theory which
copes with message length, i.e.~one which assumes that all types of
atomic messages have a specific length.  Then, reasoning about such a
theory requires arithmetic (and, in the worst case, properties of
least common multiples).  Another example extension of
Definition~\ref{def:impl} is the use of Meadow's probabilistic
approach~\cite{Meadows02} to reason about the equality of the
implementations of two messages.

\subsection{Implementation Traces}


We now introduce the new ability of the penetrator that enables it to
reinterpret a message $m$ as another message $m'$, as long as they
share a common implementation.  Since the Dolev-Yao-like abilities of
a penetrator are formalized by the set of penetrator traces (which it
can use to analyze, reuse and synthesize messages), we enlarge this
set by an additional rule, we call \emph{$I$-trace}.
\begin{definition}[Penetrator Trace]
  Let $\keys_\mathsf{P}$ denote the keys initially known to the
  penetrator. Then, a \emph{penetrator trace} of the extended strand
  space is either a penetrator strand, as given in
  Section~\ref{sec:prelim},
  or the
  following:\footnote{\onehalfspacing Recall that, for all
    $m_1,m_2\in\Terms$, $m_1=m_2$ implies $m_1\approx m_2$; though,
    clearly, this implication does not reverse.}
  \begin{tabbing}
    \hspace{0.25in}
    \=\textbf{\emph{(I)mplementation}}: $\Tuple{-m}{+m'}$, provided
    that $m\approx m'$, but $m\neq m'$.
  \end{tabbing}
  We call the messages $m$ and $m'$ \emph{camouflaged} and
  \emph{spoofed}, respectively.
\end{definition}
\begin{example}
  \label{ex:tf_WL1}
  Consider the example flawed protocol Woo Lam $\pi_1$, introduced in
  Example~\ref{ex:WooLamPi}. To capture the attack to this protocol,
  we first formalize the theory for the implementation of messages.
  So, besides the axioms given in Definition~\ref{def:impl} (which
  hold for all implementations), we require that the implementation of
  nonces can be confused with the implementation of encrypted
  messages: hence, we add the following axiom:
  \begin{equation}
    \forall z\in\Nonce. \; \exists y\in\Keys, \; x\in\Terms.\quad \imp(z) = \imp(\fat{x}_y) \label{newax}
  \end{equation}

  Then, we build the penetrator bundle describing the type flaw
  attack, as follows: 
  \begin{eqnarray*}
    \begin{array}{crlcl}
      & & \Spy & & b\\[0.2cm]
      & & +a & \xrightarrow{\hspace*{0.5in}} & -a\\
      & & \Downarrow   & & \Downarrow\\
      & & -n_b & \xleftarrow{\hspace*{0.5in}} & +n_b\\
      & \textbf{I} - trace & \Downarrow & & \Downarrow\\
      & & +\fat{m}_k & \xrightarrow{\hspace*{0.5in}} & -\fat{m}_k\\
      & & \Downarrow    & & \Downarrow\\
      & & -\fat{a\conc b\conc \fat{m}_k}_{k_{bs}} &
      \xleftarrow{\hspace*{0.5in}} & +\fat{a\conc b\conc\fat{m}_k}_{k_{bs}}\\
      & \textbf{I} - trace & \Downarrow & & \Downarrow\\
      & & +\fat{a\conc b\conc n_b}_{k_{bs}} & \xrightarrow{\hspace*{0.5in}} & -\fat{a\conc b\conc n_b}_{k_{bs}}
    \end{array} 
  \end{eqnarray*}
  This bundle contains two \textbf{I}-trace instances. The first
  instance is used to reinterpret the nonce $n_b$ as an encrypted
  message $\fat{m}_k$.  To justify the application of the trace, both
  messages have to agree on their implementation, i.e. $\imp(n_b) =
  \imp(\fat{m}_k)$. Both, $m$ and $k$, appear for the first
  time\footnote{\onehalfspacing The notion of message origination is
    an important one, and, hence, prompts adequate formalization; we
    shall do so next, see Section~\ref{sec:originates}.} at the second
  node of this \textbf{I}-trace, which basically means that the
  penetrator is free to choose $m$ and $k$ such that $\imp(n_b) =
  \imp(\fat{m}_k)$ holds.  Axiom \ref{newax} guarantees that the
  penetrator will always find appropriate values for $m$ and $k$, and
  justifies the application of the \textbf{I}-trace rule.  The second
  instance of an \textbf{I}-trace rule is used to revert the
  reinterpretation and replaces $\fat{m}_k$ by $n_b$ inside a message.
  The justification of this application is rather simple: the
  penetrator has chosen $m$ and $k$ such that $\imp(n_b) =
  \imp(\fat{m}_k)$ holds.  Applying axioms~(\ref{ax1})
  and~(\ref{ax2}), we can deduce that then also $\imp(\fat{a\conc
    b\conc n_b}_{k_{bs}}) = \imp(\fat{a\conc b\conc
    \fat{m}_k}_{k_{bs}})$ holds, which guarantees the applicability of
  the \textbf{I}-trace rule also in this case.\hfill\nqed
\end{example}

In what follows, we will abstract from this example and formalize the
justification of \textbf{I}-trace rule applications in general.

\subsection{Originating Terms and Implementation Equivalences}
\label{sec:originates}

The application of the \textbf{I}-trace rule demands that two messages
$m$ and $m'$ share the same implementation. The proof for this
precondition is done within the underlying implementation theory. In
the example above we have used the fact that the penetrator is free to
choose $m$ and $k$ appropriately. Therefore, to justify the rule
application we have to prove that:
\begin{equation}
  \forall n_b \in \Nonce. \; \exists k\in\Keys. \; \exists m\in\Terms. \quad \imp(n_b) = \imp(\fat{m}_k) \label{obl1}  
\end{equation}
which is an easy consequence of axiom (\ref{newax}). 

Notice that the proof obligation would change if the penetrator were
not free to choose $m$ and $k$, because both messages have occurred on
regular strands before applying any \textbf{I}-trace rule. In such a
case, we would have to prove:
\begin{equation}
  \label{obl2}
  \forall n_b \in \Nonce. \; \forall k\in\Keys. \; \forall m\in\Terms. \quad \imp(n_b) = \imp(\fat{m}_k)  
\end{equation}
to ensure that the penetrator can always camouflage $\fat{m}_{k}$ by
$n_b$ regardless of how $m$ and $k$ have been chosen beforehand. As a
consequence all nonces and encrypted terms would share the same
implementation. Furthermore, due to the injectivity (\ref{ax5}) of the
implementation for nonces, (\ref{obl2}) implies that the set of
messages can only contain a single nonce and a single encrypted term.

In practice, the proof obligation (\ref{obl2}) is too strong and might
prevent to reveal an attack since a honest principal could have
unintentionally chosen $m$ and $k$ in such a way that actually
$\imp(n_b) = \imp(\fat{m}_k)$ holds. However, the probability for this
to happen is similar to the probability that two principals will
independently choose the same nonce in order, for instance, to use it
as a challenge.  This actually justifies our use of (\ref{obl1}), and
also illustrates the abstraction we have made in order to avoid
probability calculations, like they are done in
\cite{Meadows02,Meadows03}.

The question of selecting either (\ref{obl1}) or (\ref{obl2}) for
justifying the first \textbf{I}-trace in
Example~\ref{ex:tf_WL1} 
depends on the fact whether $m$ and $k$ originate at the second node
of the \textbf{I}-trace or not. The notion of a node originating a
term (e.g. a nonce) comprises the principal's freedom to choose an
appropriate value for the term (e.g.  the nonce). In the original
approach, the notion of a node originating a term is a syntactical
property of strands~\cite{strand-spaces}:
\begin{quotation}
  \noindent``Let $\mathcal{B}=\Tuple{V}{(\rightarrow\cup\Rightarrow)}$
  be a bundle. An unsigned term $m$ \emph{originates} at a node $v\in
  V$, if $\sign(v)=+$; $m\sbtrm\msg(v)$; and $m\not\sbtrm\msg(v')$, for
  every $v'\Rightarrow^+ v$. $m$ is said to be \emph{uniquely
    originating} if it originates on a unique $v\in V$.''
\end{quotation}
Any information that a strand, or its respective principal learns,
occurs syntactically within some received message. Hence, a nonce
originates in a particular node of a strand if it does not occur
syntactically in any previous node of this strand.  In our setting,
the situation is different, since an information might be semantically
included in a message but camouflaged by using \textbf{I}-traces. 

For example, consider the second \textbf{I}-trace,
$\Tuple{\Spy}{4}\Rightarrow\Tuple{\Spy}{5}$, in the attack given in
Example~\ref{ex:tf_WL1}, 
Although $n_b$ occurs in
$\Tuple{\Spy}{5}$, it has been introduced in the bundle already in
$\Tuple{\Spy}{2}$, and was received by the \textbf{I}-trace
camouflaged as $\fat{m}_{k}$ in $\Tuple{\Spy}{4}$.

To guarantee that a term originates also in terms of its
implementation at a specific node, c.f.~$n_b$, we have to make sure
that it was never subject to a camouflage. Therefore, given a node $v$
of a bundle, we collect all camouflages introduced by
\textbf{I}-traces occurring in front of $v$ in the set of
\emph{equational constraints} of $v$:
\begin{definition}[Equational constraints]\label{eqctr}
  Let ${\cal B}=\Tuple{V}{(\rightarrow\cup\Rightarrow)}$ be a bundle
  and let $v\in V$. Then, the \emph{equational constraints} of $v$
  wrt.~$\mathcal{B}$, written $\Scope_{\cal B}(v)$, is a set of pairs
  given by:
  \begin{eqnarray*}
    \Scope_{\cal B}(v) & =& \{\langle\msg(v_1), \msg(v_2)\rangle \,:\, v_1\Rightarrow_{\mbox{I-trace}}v_2 \mbox{ and } v_2\preceq_\mathcal{B}v\}
  \end{eqnarray*}
  where $v_1\Rightarrow_{\mbox{I-trace}}v_2$ denotes both that $v_1$
  and $v_2$ both lie on a penetrator strand, and that they are
  interconnected via a trace of kind \textbf{I}.
\end{definition}
\begin{example}
  Going back to our running example, we find that
  $\Scope(\Tuple{\Spy}{4}) = \{\langle n_b, \fat{m}_k \rangle\}$ and
  $\Scope(\Tuple{\Spy}{5}) = \{\langle n_b, \fat{m}_k \rangle, \langle
  \fat{m}_k, n_b \rangle\}$.  Using Definition~\ref{eqctr}, we can
  adapt the notion of messages originating in a node to our
  settings.\hfill\nqed
\end{example}

\begin{definition}[Originating nodes]\label{def:origin}
  Let ${\cal B}=\Tuple{V}{(\rightarrow\cup\Rightarrow)}$ be a
  bundle and let $v\in\cal V$ be a node in $\mathcal{B}$. An atomic
  term $m$ \emph{originates} at $v$ (in $\mathcal{B}$) iff:
  \begin{itemize}
  \item $\sign(v)=+$;
  \item $m\sbtrm\msg(v)$;
  \item $m\not\sbtrm\msg(v')$ for every $v'\Rightarrow^+ v$; and
  \item $\forall \langle m', m'' \rangle \in \Scope_{\cal B}(v).\ \;
    (m\not\sbtrm m'$ and $m\not\sbtrm m'')$.
  \end{itemize}
\end{definition}
In an abuse of notation, we extend the above definition also to
encrypted messages.  Hence, an encrypted message $\fat{m'}_k$
originates at a node $v$ (in $\mathcal{B}$) iff $\fat{m'}_k$ and $v$ satisfy the
conditions of the atomic message $m$ and node $v$ in
Definition~\ref{def:origin}.  In our example of the Woo Lam $\pi_1$
protocol, both $m$ and $k$ originate at $\Tuple{\Spy}{3}$. 

	
\begin{landscape}
  \begin{figure}
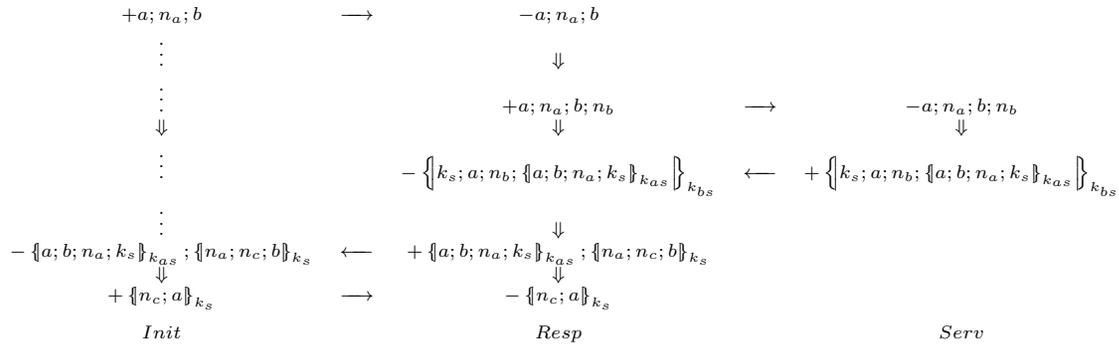

    \hrule
    \centering
    \begin{eqnarray*} 
      \scriptsize
      \begin{array}{ccccc}
        + a\conc n_a\conc b & \xrightarrow{\hspace*{0.1in}} & - a\conc n_a\conc b \\
        \vdots & & \Downarrow \\
        \vdots & & + a\conc n_a\conc b\conc n_b & \xrightarrow{\hspace*{0.1in}} & - a\conc n_a\conc b\conc n_b \\
        \Downarrow & &  \Downarrow & & \Downarrow \\
        \vdots & & - \fat{k_s\conc a\conc n_b\conc \fat{a\conc b\conc n_a\conc k_s}_{k_{as}}}_{k_{bs}} & \xleftarrow{\hspace*{0.1in}} 
        & + \fat{k_s\conc a\conc n_b\conc \fat{a\conc b\conc n_a\conc k_s}_{k_{as}}}_{k_{bs}} \\
        \vdots & & \Downarrow \\
        - \fat{a\conc b\conc n_a\conc k_s}_{k_{as}}\conc \fat{n_a\conc n_c\conc b}_{k_s} & \xleftarrow{\hspace*{0.1in}} 
        & + \fat{a\conc b\conc n_a\conc k_s}_{k_{as}}\conc \fat{n_a\conc n_c\conc b}_{k_s} \\
        \Downarrow & & \Downarrow \\
        + \fat{n_c\conc a}_{k_s} & \xrightarrow{\hspace*{0.1in}} & - \fat{n_c\conc a}_{k_s} \\[2ex]
        Init & &  Resp && Serv 
      \end{array}
    \end{eqnarray*}
    \hrule
    \caption{Bundle of the KP-protocol in~\cite{roles-crypt-prot}}
    \label{Fig:KP_bundle} 
  \end{figure}
\end{landscape}

\begin{example}[The KP Protocol]
  Another example protocol that is subject to a type flaw attack is KP
  (see Fig.~\ref{Fig:KP_bundle}).  Snekkenes~\cite{roles-crypt-prot}
  discusses this protocol and illustrates the attack.  The type flaw
  attack to KP is based on the facts that:
  \begin{enumerate}
  \item On step 3, the responder, $b$, extracts $\fat{a\conc b\conc
      n_a\conc k_s}_{k_{as}}$ from the component encrypted under
    $k_{bs}$, and, then, on step 4, sends this extracted message to $a$;
  \item Keys, agents and encrypted share common implementations
    (c.f.~the first component of the fourth message of the responder);
    and
  \item Therefore, the implementation of message 3 can be
    misinterpreted (in a different run) to be the implementation of
    the first component of message 4.
  \end{enumerate}
  Similar to the previous example, we build up an appropriate message
  theory reflecting the assumptions laid on before. Besides axioms
  (\ref{ax1})---(\ref{ax6}), we require additional axioms specifying
  the assumption that keys (generated dynamically by the server) and
  agents (or encrypted messages, respectively) potentially share the
  same implementation:
  \begin{eqnarray}
    & & \forall x\in\Agent.\;\exists w\in\Keys.\quad \imp(x) = \imp(w) \label{newax1}\\
    & & \forall w\in\Keys.\;\exists w'\in\Keys, y\in\Terms. \quad \imp(w) = \imp(\fat{y}_{w'}) \label{newax2}
  \end{eqnarray}
  Fig.~\ref{Fig:KB_penbundle} illustrates the resulting attack in the
  strand space notation using our notion of \textbf{I}-traces. This
  example penetrator bundle contains three \textbf{I}-traces. Node
  $\Tuple{\Spy}{7}$ originates the key $k'_s$ on the penetrator
  strand.  Since this key originates at $\Tuple{\Spy}{7}$, the
  penetrator is free to choose its value, and axiom (\ref{newax1})
  guarantees that there is an appropriate key $k'_s$ possessing the
  same implementation as $a$.

  Similar arguments hold for $\Tuple{\Spy}{7}$ originating $m$ and $k$
  by using axiom (\ref{newax2}).  For $\Tuple{\Spy}{9}$, it is the
  case that $\Scope_{\cal B}(\Tuple{\Spy}{9}) = \{\langle a,
  k'_s\rangle, \langle k_s, \fat{m}_k\rangle \}$, which justifies the
  application of the \textbf{I}-trace rule application, since it
  implies trivially that $\imp(\fat{m}_k) = \imp(k_s)$
  holds.\hfill\nqed
\end{example}
	
\begin{landscape}
  \begin{figure}[p]
    \hrule
    \centering
    \begin{eqnarray*}
      \begin{array}{ccccccc}
        & & & & +b\conc n'_b\conc a & \xrightarrow{\hspace*{0.2in}} & -b\conc n'_b\conc a \\
        & & & & \Downarrow & & \Downarrow \\
        & & & & -b\conc n'_b\conc a\conc n'_a & \xleftarrow{\hspace*{0.2in}}  & +b\conc n'_b\conc a\conc n'_a  \\
        & & & & \Downarrow & & \vdots \\
        & & -a\conc n'_a\conc b & \xleftarrow{\hspace*{0.2in}} & + a\conc n'_a\conc b \\
        & & \Downarrow & & \vdots & &  \\
        -\ldots & \xleftarrow{\hspace*{0.2in}} & +a\conc n'_a\conc b\conc n'_b &  \\
        \Downarrow & & \Downarrow & & \Downarrow & &  \\
        +\ldots & \xrightarrow{\hspace*{0.2in}} & -\fat{k_s\conc a\conc n_b\conc \fat{a\conc b\conc n'_a\conc k_s}_{k_{as}}}_{k_{bs}} \\
        & & \Downarrow & & \vdots & & \Downarrow \\
        & & +\fat{a\conc b\conc n'_a\conc k_s}_{k_{as}}\conc \fat{n'_a\conc n_d\conc b}_{k_s} & \xrightarrow{\hspace*{0.2in}} 
        & -\fat{a\conc b\conc n'_a\conc k_s}_{k_a}\conc \fat{n'_a\conc n_d\conc b}_{k_s} \\ 
        &  &  &      & \Downarrow & &  \\
        &  & &      & +\fat{a\conc b\conc n'_a\conc k_s}_{k_{as}} \\
        &  & \vdots &      & \Downarrow  & & \\
        &  & &      & -\fat{a\conc b\conc n'_a\conc k_s}_{k_{as}} \\
        &  &  & \textbf{I} - trace & \Downarrow & & \vdots\\
        &  & &      & +\fat{k'_s\conc b\conc n'_a\conc \fat{m}_k}_{k_{as}} & \xrightarrow{\hspace*{0.2in}}  & -\fat{k'_s\conc b\conc n'_a\conc \fat{m}_k}_{k_{as}}\\
        &  & \Downarrow &      & \Downarrow & & \Downarrow \\
        &  & &      & -\fat{m}_k\conc \fat{n'_b\conc n'_d\conc a}_{k'_s} & \xleftarrow{\hspace*{0.2in}}  & +\fat{m}_k\conc \fat{n'_b\conc n'_d\conc a}_{k'_s} \\
        &  & & \textbf{I} - trace & \Downarrow \\
        &  & & &  +k_s\conc \fat{n'_b\conc n'_d\conc a}_{k'_s} \\
        &  & \vdots &      & \Downarrow \\
        &  & -\fat{n_d\conc a}_{k_s} &   \xleftarrow{\hspace*{0.2in}}   & +\fat{n_d\conc a}_{k_s} \\[2ex]
        kdc & & b & & \Spy & & a 
      \end{array}
    \end{eqnarray*}
    \hrule
    \caption{Type flaw attack in the KP-protocol in~\cite{roles-crypt-prot}}
    \label{Fig:KB_penbundle} 
  \end{figure}
\end{landscape}

\section{Protocol Flaws}
\label{sec:principles} 

In this section, we will analyze the different types of flaws
resulting in an insecure protocol. The main source of such flaws in
security protocols is that typically a protocol is not run in
isolation, but that there are, simultaneously, multiple run instances
of the same protocol that might interfere with each other. Typically,
the penetrator carries out an attack on a protocol run, by reusing
information gathered from another run (or even from the same one).
Hence, patching security protocols is mostly concerned with
disambiguating individual protocol messages in order to avoid that a
message in some protocol run can be misinterpreted either as another
message at a different protocol step, or as a message of the same
protocol step but on a different protocol run.  We start with an
investigation of how to find the steps in a protocol that cause the
misuse of messages and then we will discuss possible changes in the
shape of such messages to avoid the confusion.

\subsection{Locating the Reuse of Messages}

In order to fix a protocol, we must analyze the input attack bundle of
the protocol with the aim of identifying the positions where messages,
eavesdropped from one or more runs, are fed into some others.  So, in
this section, we will provide the notation that is necessary to split
an attack bundle into strand sets representing the different protocol
runs involved therein.  We will also define the notion of canonical
bundle, which represents an ideal protocol run imposed by a set of
given honest strands.  Comparing the message flow in the canonical
bundle versus the attack bundle enables us to determine the crucial
steps in the protocol which cause the flawed behavior.

\subsubsection{Roles and Protocol Participation}
\label{sec:roles}

We start with the formal introduction of roles as a sort of generic
strands.  Roles are instantiated to strands, by renaming generic
atomic messages to specific elements in $\Atoms$.  We also consider
penetrator traces as roles.
\begin{definition}[Roles]
  A \emph{role} $r$ is a pair $\langle s, \mess \rangle$ of a strand $s$ and
	$\mess\subseteq\Atoms$ of atomic messages occurring in $s$.
	Intuitively, $r$ denotes a strand $s$ parameterized in $\mess$.
\end{definition}
In abuse of notation we identify a role $r$ with its strand $s$ if $\mess$
is the set of all atomic messages occurring in $s$.
\begin{definition}[$I$-th Step Execution of a Strand]
  Let $r = \langle s, \mess \rangle$ be a role and let $1\leq i\leq\depth(s)$. Then, a strand $s'$
  is an \emph{instance of $r$}, \emph{up to the execution of step
    $i$}, iff there is a renaming $\alpha$ of $\mess$, such that $s'
  = \alpha(\Tuple{s}{1}) \Rightarrow \ldots \Rightarrow
  \alpha(\Tuple{s}{i})$. We denote the strand instance $s'$ of $r$ up
  to step $i$ by $\SI{r}{i}{\alpha}$. Let
  $s'=\SI{r}{i}{\alpha}$; whenever $i=\depth(s)$, then $s'$
  is said to be \emph{complete}; otherwise, it is said to be
  \emph{partial}.

  Given a set $\mathcal{R}$ of roles, then $\mathcal{R}_H$ denotes the
  regular roles in $\mathcal{R}$, and $\mathcal{R}_P$ those of the
  penetrator.  Likewise, let $\mathcal{T}$ be a set of strand
  instances, then $\mathcal{T}_H$ denotes all regular, possibly
  partial, strand instances, and $\mathcal{T}_P$ all the penetrator
  ones.
\end{definition}
\begin{example}
  The classical Needham-Schroeder Public Key (NSPK) protocol consists
  of two roles:
  \begin{eqnarray} 
    \label{NSPRole1} 
    \mbox{initiator role } r_{\mathit{Init}} & : & 
      +\fat{a \conc n}_{k_b} \Rightarrow -\fat{n \conc n'}_{k_a}
      \Rightarrow +\fat{n'}_{k_b} 
    \\
    \label{NSPRole2} 
    \mbox{responder role } r_{\mathit{Resp}} & : & 
      -\fat{a \conc n}_{k_b} \Rightarrow +\fat{n \conc n'}_{k_a}
      \Rightarrow -\fat{n'}_{k_b} 
  \end{eqnarray}
  Then, from these roles, we may obtain the following strand
  instances:
  \begin{eqnarray*}
    \SI{r_{\mathit{Init}}}{2}{\{
        a 
        \gets a, 
        n 
        \gets n, 
        k_a 
        \gets k_a, 
        k_b 
        \gets k_b \} } & 
    : & \quad  +\fat{a \conc n}_{k_b} \Rightarrow -\fat{n \conc n'}_{k_a} \\
    \SI{r_{\mathit{Resp}}}{1}{\{
      a 
      \gets a, 
      n 
      \gets n, 
      k_b 
      \gets k_c\} } & : & \quad -\fat{a \conc n}_{k_c}
  \end{eqnarray*}
  \hfill\nqed\\[-5mm]
\end{example}

A bundle $\mathcal{B}$ is constructed from a set of roles, by
combining various role instances with the help of inter-strand
communication, $\rightarrow$.  Hence, stripping off inter-strand
communication from a bundle, $\mathcal{B}$, we end up with a multiset
$\mathcal{T}$ of isolated instances of some roles.
\begin{definition}[${\cal R}$-Bundles]
  Let $\mathcal{R}$ be a set of roles and $\mathcal{T}$ be a multiset
  of (possibly partial) strand instances of roles in $\mathcal{R}$.
  Then, $\mathcal{B}$ is a $\mathcal{R}$-\emph{bundle out of}
  $\mathcal{T}$ iff (i) each node in a strand of $\mathcal{T}$ is also
  a node in $\mathcal{B}$ and vice versa, and (ii) $n \Rightarrow n'$
  iff $n \Rightarrow_{\mathcal{B}} n'$ for all nodes $n,n'$ of
  $\mathcal{B}$ (and $\mathcal{T}$ respectively).
\end{definition}

We define a canonical bundle to be a bundle representing an intended
run of a protocol. In a canonical bundle, each role is instantiated by
a strand exactly once, and each strand instance is complete.
Furthermore, the strands of a canonical bundle, and thereby the
associated roles, are all regular.
\begin{definition}[Canonical Bundles]
  An $\mathcal{R}$-bundle $\mathcal{B}$ out of $\mathcal{T}$ is
  \emph{canonical} iff all roles in $\mathcal{R}$ are regular and
  there is a symbol renaming $\alpha$ that is injective wrt.\ each
  role $r \in \mathcal{R}$ such that $\mathcal{T} = \{
  \SI{r}{\depth(r)}{\alpha} \;|\; r \in \mathcal{R} \}$
\end{definition}
Given a canonical $\mathcal{R}$-bundle $\mathcal{B}$ we can easily
retrieve $\mathcal{R}$ (up to isomorphism) from $\mathcal{B}$ (or
$\mathcal{T}$) by collecting all (renamed) strands occurring in
$\mathcal{B}$ (or $\mathcal{T}$, respectively).

Furthermore, we can consider each finite, acyclic graph
$\mathcal{B}=\Tuple{V}{(\rightarrow\cup\Rightarrow)}$ as a canonical
$\mathcal{R}$-bundle if for all $v_2\in V$ $\sign(v_2)=-$ implies that
there is a unique $v_1\in V$ with $v_1\rightarrow v_2$, and that
$\mathcal{R}$ is the set of all roles extracted from $\mathcal{B}$.
Later on, we use this property in order to formulate changes of a
protocol (i.e. changes of the roles of a protocol) simply by changing
the original canonical protocol and interpret the resulting graph as
the canonical bundle of the new protocol comprising also its new roles
(up to isomorphism).

In what follows, we will use sans serif fonts to indicate messages in
canonical bundles. Furthermore, we use the superscript $c$ to denote a
canonical bundle $\mathcal{B}^c$, and the roles $\mathcal{R}^c$ or set
of strand instances $\mathcal{T}^c$ of a \emph{canonical} bundle.  In
the examples, we also assume that the honest roles of a canonical
bundle are normalized so that $\alpha = \{\}$, and then write
$\Tuple{r}{i}$, instead of $\Tuple{\SI{r}{\depth(r)}{\{\}}}{i}$.  To
illustrate our ideas and definitions, we will make use of various
protocols as running examples.  We first present canonical bundles of
these protocols.
\begin{example}
  \label{Ex:NSPK_bundle}
  We start with the NSPK protocol. Given the roles
  $\mathcal{R}^{c}_{\mathrm{NSPK}} = \{\mathsf{Init}, \mathsf{Resp}\}$ of NSPK, (\ref{NSPRole1}) and
  (\ref{NSPRole2}), we obtain the following canonical bundle,
  $\mathcal{B}^c$:
{\sansmath
    \begin{eqnarray*}  
      \begin{array}{ccc}
        +\fat{a \conc n}_{k_b} & \xrightarrow{\hspace*{0.5in}} & -\fat{a \conc n}_{k_b}\\
        \Downarrow & & \Downarrow\\
        -\fat{n \conc n'}_{k_a}  & \xleftarrow{\hspace*{0.5in}} & +\fat{n \conc n'}_{k_a} \\
        \Downarrow & & \Downarrow\\
        +\fat{n'}_{k_b} & \xrightarrow{\hspace*{0.5in}} & -\fat{n'}_{k_b} \\[2ex]
        Init && Resp
      \end{array} 
    \end{eqnarray*}}
  \hfill\nqed
\end{example}

\begin{example}\label{Ex:WMF_bundle}
  As another example, consider the Wide-Mouth-Frog (WMF) protocol. 
	With $\mathcal{R}^{c}_{\mathrm{WMF}} = \{\mathsf{Init}, \mathsf{Serv}, \mathsf{Resp}\}$, the
  canonical bundle of this protocol is as follows:
{\sansmath
    \begin{eqnarray*}
      \begin{array}{ccccc}
        + a, \fat{b \conc t_a \conc k}_{k_{as}} &   \xrightarrow{\hspace*{0.1in}} &  - a, \fat{b \conc t_a \conc k}_{k_{as}} \\    
        & &    \Downarrow  \\
        & & + \fat{a \conc t_{a+d} \conc k}_{k_{bs}} &   \xrightarrow{\hspace*{0.1in}} & - \fat{a \conc t_{a+d} \conc k}_{k_{bs}} \\[2ex]
        Init && Serv && Resp
      \end{array}
    \end{eqnarray*}}
    \label{ex:WMF_bundle}
    \hfill\nqed
\end{example}

\begin{example}\label{Ex:WLAM_bundle}
  As a third example, look into the Woo Lam authentication protocol:
{\sansmath
    \begin{eqnarray*}   \scriptsize
      \begin{array}{ccccc}
        +a \conc n  & \xrightarrow{\hspace*{0.1in}} &  - a \conc n \\  
        \Downarrow   &  &  \Downarrow \\
        -b \conc n'  & \xleftarrow{\hspace*{0.1in}} &  + b \conc n' \\  
        \Downarrow   &  &  \Downarrow \\
        + \fat{a \conc b \conc n \conc n'}_{k_{as}} & \xrightarrow{\hspace*{0.1in}} & - \fat{a \conc b \conc n \conc n'}_{k_{as}} \\
        & &  \Downarrow \\
        && + \fat{a \conc b \conc n \conc n'}_{k_{as}}	\conc \fat{a \conc b \conc n \conc n'}_{k_{bs}}	& \xrightarrow{\hspace*{0.1in}}  
        & -	\fat{a \conc b \conc n \conc n'}_{k_{as}}	\conc \fat{a \conc b \conc n \conc n'}_{k_{bs}}	\\
        &&  \Downarrow & & \Downarrow \\														
        && - \fat{b \conc n \conc n' \conc k_{ab}}_{k_{as}} \conc \fat{a \conc n \conc n' \conc k_{ab}}_{k_{bs}}		
        & \xleftarrow{\hspace*{0.1in}}  & + \fat{b \conc n \conc n' \conc k_{ab}}_{k_{as}} \conc \fat{a \conc n \conc n' \conc k_{ab}}_{k_{bs}} \\
        & & \Downarrow \\
        - \fat{b \conc n \conc n' \conc k_{ab}}_{k_{as}} \conc \fat{n \conc n'}_{k_{ab}} 
        & \xleftarrow{\hspace*{0.1in}}  & + \fat{b \conc n \conc n' \conc k_{ab}}_{k_{as}} \conc \fat{n \conc n'}_{k_{ab}} \\[2ex]						
        Init && Resp && Serv   	
      \end{array} 
    \end{eqnarray*}}
    \hfill\nqed
\end{example}

\begin{example}\label{Ex:DSSK_bundle}
  Our last example is the Denning-Sacco-Shared Key (DSSK) protocol.  {
    \sansmath
    \begin{eqnarray*} \scriptsize
      \begin{array}{ccccc}
        & & + a \conc b  & \xrightarrow{\hspace*{0.1in}} & - a \conc b  \\
        & & \Downarrow   & & \Downarrow \\
        & & - \fat{b \conc k_{ab} \conc t_s \conc \fat{b \conc k_{ab} \conc a \conc t_s}_{k_{bs}}}_{k_{as}} & \xleftarrow{\hspace*{0.1in}} &
        + \fat{b \conc k_{ab} \conc t_s \conc \fat{b \conc k_{ab} \conc a \conc t_s}_{k_{bs}}}_{k_{as}}  \\
        & & \Downarrow  \\
        -\fat{b \conc k_{ab} \conc a \conc t_s}_{k_{bs}} & \xleftarrow{\hspace*{0.1in}}  & +\fat{b \conc k_{ab} \conc a \conc t_s}_{k_{bs}} \\
        Resp && Init && Serv   	
      \end{array} 
    \end{eqnarray*}}
  \hfill\nqed
\end{example}

\subsubsection{Characterizing the Perspective of each Participant in a Protocol Attack}

Attacks detected by some security protocol analyzer typically involve
different (partial) protocol runs. Messages generated in one run are
misused to fake messages in other runs. Since honest principals only
see the messages of a protocol that they receive or send, each
principal develops his own view on the protocol run. In the attack of
the NSPK protocol, see Example~\ref{ex:bundle}, 
Alice assumes she is communicating with Charly, while Bob thinks he is
communicating with Alice. However, both views are not compatible. In
the following definition, we truss compatible strands of honest
principals to so-called protocol sections.  While in a canonical
bundle all honest strands belong to a single protocol section, the
attack bundle of NSPK consists of two protocol sections denoting the
(faked) protocol runs of Alice with Charly and Alice with Bob,
respectively.
\begin{definition}[Protocol Section]
  \label{def:section}
  Let $\mathcal{B}^c$ be a canonical $\mathcal{R}^c$-bundle out of some
  $\mathcal{T}^c$ and let $\mathcal{B}$ be an $\mathcal{R}$-bundle out
  of some $\mathcal{T}$. 

  A set $\mathcal{T}' \subseteq \mathcal{T}$ is a \emph{protocol
    section} wrt.~$\mathcal{B}^c$ iff there is a renaming $\beta$ such
  that, for all $\SI{r}{i}{\alpha'} \in \mathcal{T}'$, there is one,
  and only one, $\SI{r}{\depth(r)}{\alpha} \in \mathcal{T}^c$, with
  $\alpha'$ being a function denoting the composition of $\beta$ and
  $\alpha$ and subject to the domain $\Atoms(r)$.
  We call $\beta$ the \emph{canonical renaming} for $\mathcal{T}'$
  wrt.~$\mathcal{B}^c$.
	
  We call the mapping that maps each node of a regular strand
  $\Tuple{\SI{r}{i}{\alpha'}}{j}$ in $\mathcal{T}'$ to the node
  $\Tuple{\SI{r}{i}{\alpha}}{j}$ in $\mathcal{B}^c$, for all 
	$\SI{r}{i}{\alpha'} \in \mathcal{T}'$ and $1 \leq j \leq i$, the
  \emph{canonical mapping} of $\mathcal{T}'$ to $\mathcal{B}^c$.
\end{definition}

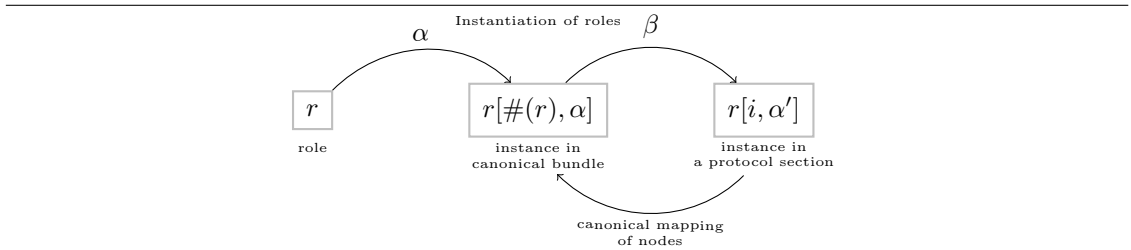
\begin{figure}[htb]
  \begin{center}
    \hrule
    \begin{tikzpicture}
      [place/.style={rectangle,draw=gray!50,thick, inner sep=5pt,minimum size=4mm}]
      \node at (0, 2.2) {\tiny Instantiation of roles};
      \node at (-3, 1) [place] (n1) {$r$};
      \node at (0, 1)  [place] (n2) {$r[\#(r), \alpha]$};
      \node at (3, 1)  [place] (n3) {$r[i, \alpha']$};
      \node at (-3, 0.5) {\tiny role};
      {\tiny \node at (0, 0.4) [text width=2cm, text centered]  (n4) {instance in \\ canonical bundle}; 
        \node at (3, 0.4) [text width=2cm, text centered]  (n5) {instance in \\ a protocol section}; }
      \draw (n1)[->] to [out=45,in=135] node [sloped,above] {$\alpha$} (n2) ;
      \draw (n2)[->] to [out=45,in=135] node [sloped,above] {$\beta$}  (n3);
      {\tiny  \draw (n5)[->] to [out=-135,in=-45] node [sloped,below, midway, text width=2cm, text centered] {canonical mapping \\ of nodes}  (n4);}
    \end{tikzpicture}
    \hrule
    \caption{Relations of role instances} 
    \label{fig:rolerelation}
  \end{center}
\end{figure}

Fig.~\ref{fig:rolerelation} illustrates the relations between roles
and their instances in the canonical bundle and a protocol section,
respectively.
\begin{example}
  \label{ex:bundle}
  To exemplify all these notions, let us study again the attack on
  NSPK, illustrated in the bundle shown below.
  \begin{eqnarray*}
    \begin{array}{ccccccccc}
      &   &   -k_c^{-1}       & \xleftarrow{\hspace*{0.1in}} &   + k_c^{-1} \\    
      &   &    \Downarrow  \\
      +\fat{a \conc n}_{k_c} &  \xrightarrow{\hspace*{0.1in}}  & -\fat{a \conc n}_{k_c} &  &    +k_b  &  \xrightarrow{\hspace*{0.1in}} & -k_b     \\
      \vdots &   &    \Downarrow     &  &              &  &  \Downarrow \\
      &   &   + a \conc n          &  \multicolumn{3}{c}{\xrightarrow{\hspace*{0.7in}}}  & -a \conc n       \\
      \Downarrow 	  		&   &                   &  &              &  &   \Downarrow  \\
      \vdots &   &                   &  &              &  &  + \fat{a \conc n}_{k_b}  & \xrightarrow{\hspace*{0.1in}}  & -\fat{a \conc n}_{k_b}\\
      &   &  &&  && &&  \Downarrow \\
      -\fat{n \conc n'}_{k_a}	& \multicolumn{7}{c}{\xleftarrow{\hspace*{2.3in}}} & +\fat{n \conc n'}_{k_a} \\
      \Downarrow  	&   &  &&  && &&   \vdots  \\			
      &    &  -k_c^{-1} & \xleftarrow{\hspace*{0.1in}} & +k_c^{-1} \\
      &    &  \Downarrow \\
      +\fat{n'}_{k_c}  &  \xrightarrow{\hspace*{0.1in}} & -\fat{n'}_{k_c} & & +k_b & \xrightarrow{\hspace*{0.1in}} & -k_b & & \Downarrow\\
      &    & \Downarrow & & & &  \Downarrow \\ 
      &    &   +n'   & \multicolumn{3}{c}{\xrightarrow{\hspace*{0.5in}}} &  -n' & &   \vdots \\
      &   &   & & & & \Downarrow \\
      &   &   & & & & +\fat{n'}_{k_b} & \xrightarrow{\hspace*{0.1in}} & -\fat{n'}_{k_b}\\[2mm]
      \emph{Alice} &&&&\Spy &&&& \emph{Bob}
    \end{array} 
  \end{eqnarray*}
  Besides the two roles of $\mathcal{R}_{\mathrm{NSPK}}$, this bundle
  contains the penetrator traces \textbf{(K)}, \textbf{(E)}, and
  \textbf{(D)}.  It is therefore an $\mathcal{R}$-bundle, with
  $\mathcal{R} = \{ \textbf{(K)}, \textbf{(E)}, \textbf{(D)} \} \cup
  \mathcal{R}_{\mathrm{NSPK}}$.  $\mathcal{T}$ consists of the six
  individual strands occurring in the bundle.
 
  Using the canonical bundle $\mathcal{B}^c$ given in
  Example~\ref{Ex:NSPK_bundle}, this attack bundle contains an
  instance $\SI{r_{\mathit{init}}}{3}{\beta}$ with $\beta =
  \{\mathsf{a} \gets a, \mathsf{b} \gets c,\mathsf{n} \gets n,
  \mathsf{n'} \gets n', \mathsf{k_a} \gets k_a, \mathsf{k_b} \gets k_c
  \}$ and an instance $\SI{r_{\mathit{resp}}}{3}{\beta'}$ with $\beta'
  = \{\mathsf{a} \gets a, \mathsf{b} \gets b,\mathsf{n} \gets n,
  \mathsf{n'} \gets n', \mathsf{k_a} \gets k_a, \mathsf{k_b} \gets k_b
  \}$.  Both form individual protocol sections $\mathcal{T}_1 = \{
  \SI{r_{\mathit{init}}}{3}{\beta} \}$ and $\mathcal{T}_2 = \{
  \SI{r_{\mathit{resp}}}{3}{\beta'} \}$.  However, $\mathcal{T}_3 =
  \mathcal{T}_1 \cup \mathcal{T}_2$ is not a protocol section, because
  $\beta$ and $\beta'$ are incompatible. \hfill\nqed
\end{example}

\begin{example}\label{Ex:WMF_penbundle}
  Concerning the WMF protocol (see Example~\ref{Ex:WMF_bundle})
  AVISPA~\cite{avispa} has found that it fails to guarantee weak
  authentication of the responder, Bob, to the initiator, Alice,
  yielding the attack bundle shown below: 
  \begin{eqnarray*}
    \begin{array}{ccccccc}
      + a, \fat{b \conc t_a \conc k}_{k_{as}} &   \xrightarrow{\hspace*{0.1in}} &  - a, \fat{b \conc t_a \conc k}_{k_{as}} \\    
      & &    \Downarrow  \\
      & &    + a & \xrightarrow{\hspace*{0.1in}} & -a \\
      & &    \Downarrow  \\
      & & + \fat{b \conc t_a \conc k}_{k_{as}}  & \multicolumn{3}{c}{\xrightarrow{\hspace*{0.5in}}} & - \fat{b \conc t_a \conc k}_{k_{as}} \\[2ex]
      Alice (Init) && \Spy &&&& Alice (Resp)
    \end{array}
  \end{eqnarray*}

  Comparing this attack bundle against the canonical one,
  $\mathcal{B}^c$, shown in Example~\ref{ex:WMF_bundle}, we notice
  that it contains instances for the initiator and responder strands,
  while the server is impersonated by the penetrator:
  $\SI{r_{\mathit{init}}}{1}{\beta}$ with
  $\beta = \{\mathsf{a} \gets a, \mathsf{b} \gets b,
  \mathsf{t_a} \gets t_a, \mathsf{k_{as}} \gets k_{as} \}$ and
  $\SI{r_{\mathit{resp}}}{1}{\beta'}$ with
  $\beta' = \{\mathsf{a} \gets b, \mathsf{b} \gets a,
  \mathsf{t_a} \gets t_a, \mathsf{k_{bs}} \gets k_{as} \}$. Both
  strands form individual protocol sections, $\mathcal{T}_1 = \{
  \SI{r_{\mathit{init}}}{1}{\beta} \}$ and
  $\mathcal{T}_2 = \{ \SI{r_{\mathit{resp}}}{1}{\beta'}
  \}$. Again, $\mathcal{T}_1\cup\mathcal{T}_2$ is not a protocol
  section, 
  because $\beta_{\mathit{init}}$ and $\beta_{\mathit{resp}}$ are not
  compatible.\hfill\nqed
\end{example}

\begin{example}\label{Ex:DSSK_penbundle}
  DSSK is vulnerable to a so-called multiplicity attack (see
  Section~\ref{sec:p9}):

  \begin{eqnarray*}  \sansmath \scriptsize
    \begin{array}{ccccccc}
		&& & & + a \conc b  & \xrightarrow{\hspace*{0.1in}} & - a \conc b  \\
		&& & & \Downarrow   & & \Downarrow \\
		&& & & - \fat{b \conc k_{ab} \conc t_s \conc \fat{b \conc k_{ab} \conc a \conc t_s}_{k_{bs}}}_{k_{as}} & \xleftarrow{\hspace*{0.1in}} &
		    + \fat{b \conc k_{ab} \conc t_s \conc \fat{b \conc k_{ab} \conc a \conc t_s}_{k_{bs}}}_{k_{as}}  \\
		&& & & \Downarrow  \\
		& & - \fat{b \conc k_{ab} \conc a \conc t_s}_{k_{bs}} & \xleftarrow{\hspace*{0.1in}}  & + \fat{b \conc k_{ab} \conc a \conc t_s}_{k_{bs}} \\
		&& \Downarrow \\
		- \fat{b \conc k_{ab} \conc a \conc t_s}_{k_{bs}} & \xleftarrow{\hspace*{0.1in}}  & + \fat{b \conc k_{ab} \conc a \conc t_s}_{k_{bs}} \\
		&& \Downarrow \\
		- \fat{b \conc k_{ab} \conc a \conc t_s}_{k_{bs}} & \xleftarrow{\hspace*{0.1in}}  & + \fat{b \conc k_{ab} \conc a \conc t_s}_{k_{bs}} \\
	        Resp && Spy && Init && Serv   	
   \end{array} 
  \end{eqnarray*}
  \hfill\nqed
\end{example}

Thus, apart from penetrator activities, an attack bundle contains
honest strands that operate under different assumptions (captured by
the specific elements in $\Atoms$ instantiating each strand).  This is
in contrast with a canonical bundle, where all strands operate for the
same run (parameters). A protocol section hence groups together all
the honest strands in the attack bundle operating under the same
assumptions. This is the rationale that enables us to compute the
minimal set of protocol sections which, together with the penetrator's
activities, constitutes a given attack bundle.
\begin{definition}[Coverage]
  Let $\mathcal{B}$ be an $\mathcal{R}$-bundle out of some
  $\mathcal{T}$ and let $\mathcal{B}^c$ be a canonical
  $\mathcal{R}^c$-bundle 
  with $\mathcal{R}_H \subseteq \mathcal{R}^c$.
  A partition $\mathcal{T}_1,\ldots, \mathcal{T}_k$ of $\mathcal{T}_H$
  is a \emph{coverage} of $\mathcal{T}$ wrt.~$\mathcal{B}^c$ iff each
  $\mathcal{T}_i$ is a protocol section wrt.~$\mathcal{B}^c$. A
  coverage is \emph{optimal} iff it is not a
  refinement 
  of any other coverage. The canonical mapping of $\mathcal{B}$ to
  $\mathcal{B}^c$ is the composition of the canonical mappings of
  $\mathcal{T}_1,\ldots, \mathcal{T}_k$ to $\mathcal{B}^c$.
\end{definition}
\begin{example}
  Let us illustrate coverage through two examples.  Go back to
  Example~\ref{ex:bundle}; there, $\{\mathcal{T}_1, \mathcal{T}_2 \}$
  is a coverage of $\mathcal{T}$; it is also optimal since
  $\mathcal{T}_3$ is not a protocol section.  Now go back to the
  attack bundle given in Example~\ref{Ex:DSSK_penbundle}; then, we
  obtain an optimal coverage consisting of two protocol sections: the
  first covers an instance of the entire canonical bundle, while the
  second consists only of the second copy of the responder
  strand.\hfill\nqed
\end{example}

\subsubsection{The Misuse of Messages}

In what follows, we characterize situations in which the penetrator
confuses an honest principal, by misusing an observed encrypted
message. In a first step, we characterize the conditions under which a
honest principal will accept a received message, because it satisfies
his expectations, as formulated in the definition of the protocol.
\begin{definition}[Acceptable Messages]
  Let $m, m'$ be messages, $\keys$ a set of keys and $Vars \subset
  \Atoms(m)$ be a set of atomic messages in $m$. Then, $m'$ \emph{is
    accepted for} $m$ wrt.~a renaming $\sigma$ with $DOM(\sigma)
  \subseteq Vars$ and keys $\keys$, iff $\vdash_{\keys} (m, m',
  \sigma)$ according to the following rules:
  \begin{flalign}
    \tag{Id} \frac{}{\vdash_{\keys}(m,m,\langle\rangle)} & \quad & \label{rule:id} \\ 
    \tag{Sub} 
		\frac{}{\vdash_{\keys}(m, m', \sigma) } && 
    \begin{minipage}{.4\textwidth} 
      if $Dom(\sigma) \subseteq Vars \cap \Atoms(m)$
      $\imp(\sigma(m)) = \imp(m')$,  and
      $ m\, \in\, Vars$ or  $m = \fat{m''}_{k} \mbox{ with } k^{-1} \not\in K$
    \end{minipage} \label{rule:sub}  \\[0.2cm]
    \frac{ \vdash_{\keys}(m_1, m'_1, \sigma_1) \quad \vdash_{\keys}(m_2, m'_2, \sigma_2)} 
    { \vdash_{\keys}(m_1 \conc m_2, m', \sigma_1 \circ \sigma_2) } 
    && \begin{minipage}{0.4\textwidth}
		   \mbox{if } $\sigma_1, \sigma_2$ compatible and \\ $\imp(m') = \imp(m'_1 \conc m'_2)$
		   \end{minipage} \tag{Seq} \label{rule:seq} \\[0.2cm]
    \frac{ \vdash_{\keys}(k, m_2, \sigma_2) \quad \vdash_{\keys}(m, m_1, \sigma_1)}
    { \vdash_{\keys}(\fat{m}_{k}, m', \sigma_1\circ \sigma_2) } 
    &&
    \begin{minipage}{0.4\textwidth}
      \mbox{if } $\sigma_1,\sigma_2\mbox{ compatible, }\text{ and }$\\
      $\imp(m') = \imp(\fat{m_1}_{m_2})$ with $k^{-1} \in K$
    \end{minipage} \tag{Enc} \label{rule:enc} 
  \end{flalign}
	We say $m'$ \emph{is accepted for} $m$ wrt.\ atoms $Vars$ and keys $\keys$
	if there is a renaming $\sigma$ with $DOM(\sigma) \subseteq Vars$ such
	that $m'$ \emph{is accepted for} $m$ wrt.\ $\sigma$ and $\keys$.
\end{definition}
\begin{example}
  \label{ex:nslpk}
  For example, consider again the attack to the NSPK protocol (see
  Example~\ref{ex:bundle}). 
  For the sake of simplicity, assume that $\imp$ is the identity
  function.  In the second step of her strand, Alice, $a$, expects a
  message $m = \fat{n_a\conc n_b}_{k_a}$.  Suppose she receives $m' =
  \fat{n_a\conc n_c}_{k_a}$, then we can derive:
  \[\infer[\ref{rule:enc}]{\hspace*{0.5in}\vdash_{\keys}(\fat{n_a \conc n_b}_{k_a},
    \fat{n_a \conc n_c}_{k_a}, \langle n_b \gets n_c \rangle)\hspace*{0.5in}}
     {\infer[\ref{rule:id}]{\vdash_{\keys}(k_a,k_a,\langle\rangle)}{-}
	  \quad\infer[\ref{rule:seq}]{\hspace*{0.25in}\vdash_{\keys}(n_a \conc n_b, n_a \conc n_c, \langle n_b \gets n_c \rangle)\hspace*{0.5in}}
    {\infer[\ref{rule:id}]{\vdash_{\keys}(n_a, n_a, \langle \rangle)}{-} &
      \infer[\ref{rule:sub}]{\vdash_{\keys}(n_b, n_c, \langle n_b \gets n_c \rangle)}{-}}} \]
  \hfill\nqed
\end{example}
\begin{example}
  As a second example, consider now the type flaw attack on the flawed
  Woo Lam $\pi_1$ protocol, given in Example~\ref{ex:tf_WL1} (see
  Example~\ref{ex:tf_WL1}). 
  In step 3 of his strand, Bob, $b$, expects a message of the form
  $\fat{a \conc b \conc n_b}_{k_{as}}$, but receives $n_b$ instead.
  According to our definition, $b$ will accept $n_b$ provided that
  there is an instantiation $\sigma$ for $n_b$ and $k_{as}$ such that
  $\imp(\sigma(\fat{a \conc b \conc n_b}_{k_{as}})) = \imp(n_b)$.
  Assuming $b$ takes the key $k_{as}$ to be $k$, then we would infer
  the following conditions as those under which $b$ accepts $n_b$ as a
  legal message:
  \[\infer[\ref{rule:enc}]{\hspace*{1.25in}\vdash_{\keys}(\fat{a\conc b\conc n_b}_{k}, n_b,\langle n_b\gets m_3'\rangle)\hspace*{1.25in}}
	  {\infer[\ref{rule:id}]{\vdash_{\keys}(k,k,\langle\rangle)}{-}\quad\quad
      \infer[\ref{rule:enc}]{\vdash_{\keys}(a\conc b\conc n_b, a\conc b\conc m_3',\langle n_b\gets m_3'\rangle))}
         {\infer[\ref{rule:seq}]{\vdash_{\keys}(a\conc b, a\conc b,\langle\rangle)}{
          \infer[\ref{rule:id}]{\vdash_{\keys}(a, a,\langle \rangle )}{-}\quad\quad
          \infer[\ref{rule:id}]{\vdash_{\keys}(b, b,\langle \rangle )}{-}}
        \quad\quad
        \infer[\ref{rule:seq}]{\vdash_{\keys}(n_b, m_3',\langle n_b\gets m_3'\rangle)}{-}}}
    \]
    which holds, provided that there are messages $m', m_1', m_2',
    m_3'$ such that $\imp(a) = \imp(m_1')$, $\imp(b) = \imp(m_2')$,
    and $\imp(m') = \imp(m_1' \conc m_2' \conc m_3')$,
    $\imp(\fat{m'}_k) = \imp(\fat{a\conc b\conc m_3'}_{k})$ and
    $\imp(n_b) = \imp(\fat{m'}_k)$ holds.  \hfill\nqed
\end{example}

Now, an obvious misuse is that the penetrator sees an encrypted
message in a protocol run and reuses it in another run, but in the
same step. We can easily detect this case, by comparing the protocol
sections of the node where the message is originating and the node
where the message is received by the honest principal. We call such a
situation a cross-protocol-confusion. Another misuse is to use some
encrypted message observed in a run to camouflage another message of a
different step (but possibly in the same run), because they coincide
in their structure. In order to detect this new situation, we compare
the attack bundle with the canonical bundle and check whether the
encrypted message in the attack bundle and its counterpart in the
canonical bundle originate at corresponding message positions. We call
this alternative situation message confusion. 

The definition below captures our intuition of the misuse of messages;
it makes use of the standard notion of \emph{message position}, which
is an address represented by a string. Roughly, the subterm of $m$, at
position $\pi$, $m|\pi$, is captured as follows: $m|\epsilon=m$,
$f(m_1,\ldots,m_k)|i\cdot p=m_i|p$, where $\epsilon$ and $\cdot$
respectively stand for the empty string, and string concatenation. We
use $m[\pi\leftarrow m']$ to denote the message that results from
replacing $m|\pi$ with $m'$ in $m$.
\begin{definition}[Confusion]\label{def:confusion}
  Let $\mathcal{B}$ be an $\mathcal{R}$-bundle out of some
  $\mathcal{T}$, $\mathcal{B}^c$ be a canonical $\mathcal{R}^c$-bundle
  with $\mathcal{R}_H \subseteq \mathcal{R}^c$.  Let $\mathcal{C}$ be
  an optimal coverage of $\mathcal{T}$ wrt.~$\mathcal{B}^c$ and
  $\theta$ be the corresponding canonical mapping.  Then, there is a
  \emph{confusion} in a honest node $v \in \mathcal{B}$ iff there is
  a position $\pi$ of $v$ with $\fat{m}_k = \msg(v)|\pi$, originating in an
  honest node $v'\in\mathcal{B}$, at some position $\pi'$ of
  $\msg(v')$, such that either:
  \begin{itemize}
  \item $v$ and $v'$ are located on strands of different protocol
    sections, giving rise to a \emph{cross-protocol confusion} and
    \emph{$v\in{\cal B}$ \emph{causing} the cross-protocol confusion 
		with $\fat{m}_k$}, or
  \item 
		$\msg(\theta(v))|\pi$ does not originate in $\theta(v')$ or
		$\Scope_{\mathcal{B}^c}(\theta(v')) \not\,\vdash
    \imp(\msg(\theta(v')) | \pi') = \imp(\msg(\theta(v))|\pi)$,
		giving rise to a 
		\emph{message confusion} and \emph{$v\in{\cal B}$ \emph{causing} the 
		message confusion with $\fat{m}_k$}.
  \end{itemize}
\end{definition} 
\begin{example}
  In the NSPK attack bundle (see 
  Example~\ref{ex:bundle}), there is only one node that gives rise to
  a confusion. While both $\fat{a \conc n}_{k_b}$ and $\fat{n'}_{k_b}$
  originate on a penetrator strand, $\fat{n, n'}_{k_a}$ originates on
  $\Tuple{\SI{r_{\mathit{resp}}}{3}{\beta'}}{2}$.  This latter message
  is later received in $\Tuple{\SI{r_{\mathit{init}}}{3}{\beta}}{2}$,
  which belongs to a different protocol section. Hence, there is a
  cross-protocol confusion in
  $\Tuple{\SI{r_{\mathit{init}}}{3}{\beta}}{2}$.  There is not a
  message confusion, because $\fat{n, n'}_{k_a}$ originates in
  $\theta(\Tuple{\SI{r_{\mathit{resp}}}{3}{\beta'}}{2}) =
  \Tuple{r_{\mathit{resp}}}{2}$ of the canonical bundle at position
  root. Notice also that $\emptyset \vdash \fat{n, n'}_{k_a} = \fat{n,
    n'}_{k_a}$ holds.\hfill\nqed
\end{example}
\begin{example}			
  In the Woo Lam $\pi_1$ attack bundle (see
  Example~\ref{ex:tf_WL1}), 
  the principal $b$ receives an encrypted message in
  $\Tuple{\SI{r_{\mathit{resp}}}{5}{\beta}}{5}$, which originates in
  $\Tuple{\SI{r_{\mathit{resp}}}{5}{\beta}}{4}$.  This is different to
  the canonical bundle, where $\msg(\Tuple{r_{\mathit{resp}}}{5})$
  originates at $\Tuple{r_{\mathit{serv}}}{2}$. Thus, there is a
  message confusion in
  $\Tuple{\SI{r_{\mathit{resp}}}{5}{\beta}}{5}$.\hfill\nqed
\end{example}
\begin{example}	
  In the WMF protocol attack bundle (see
  Example~\ref{Ex:WMF_penbundle}), 
  there is both a message confusion and a cross-protocol confusion in
  $\Tuple{\SI{r_{\mathit{resp}}}{1}{\beta'}}{1}$.\hfill\nqed
\end{example}

\begin{landscape}
  \begin{figure}[p]
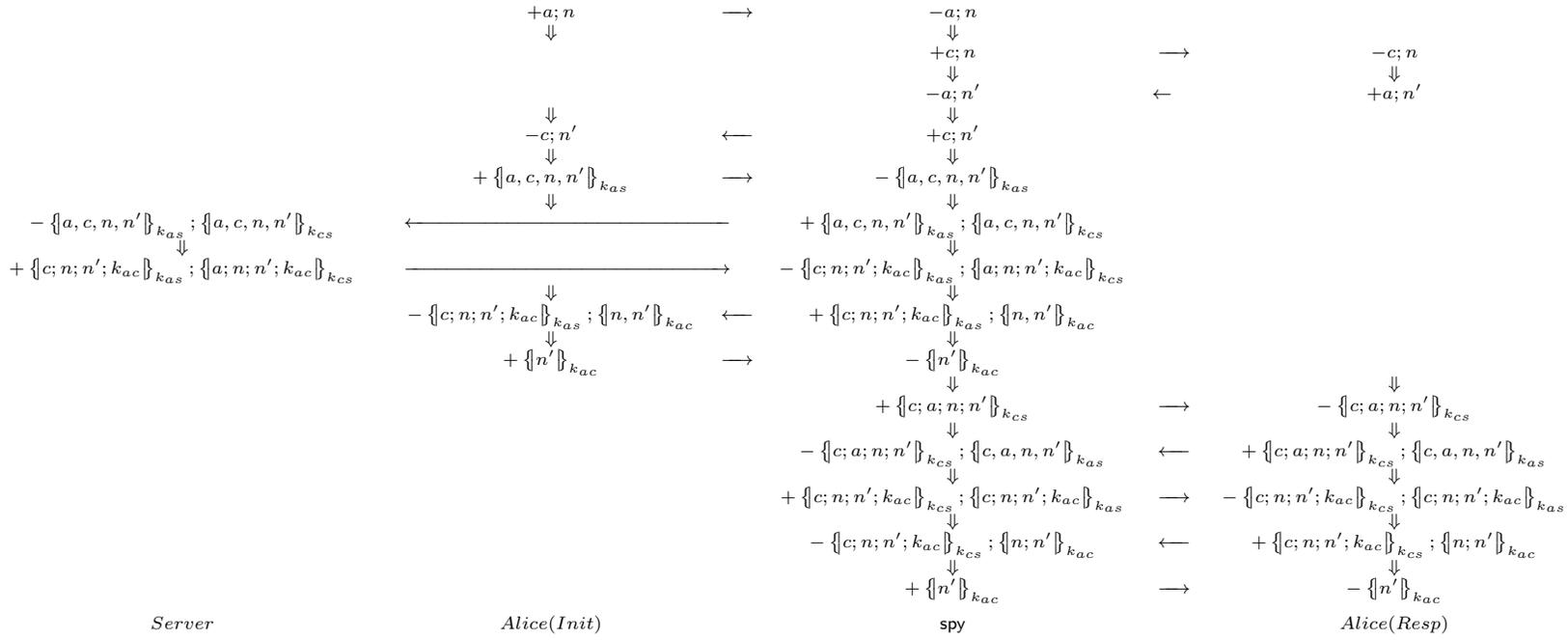
\hrule
    \begin{eqnarray*}  \scriptsize
      \begin{array}{ccccccc}
   & &      +a \conc n  & \xrightarrow{\hspace*{0.1in}} &  - a \conc n  \\  
   & &          \Downarrow   &  &  \Downarrow \\
	& & 			& &  + c \conc n & \xrightarrow{\hspace*{0.1in}} & -c \conc n 	\\
	& & 			& &  \Downarrow  &                               & \Downarrow \\
	& & 			& &  -a \conc n' & \leftarrow{\hspace*{0.1in}} &  +a \conc n' \\
	& & 		\Downarrow  	& &  \Downarrow  \\
	& & 		 - c \conc n' & \xleftarrow{\hspace*{0.1in}} & + c \conc n'  \\
	& & 		\Downarrow  &  &  \Downarrow  \\   
	& & 			+ \fat{a, c, n, n'}_{k_{as}} & \xrightarrow{\hspace*{0.1in}} & - \fat{a, c, n, n'}_{k_{as}} \\
	& &     \Downarrow  &  &  \Downarrow  \\  
	- \fat{a, c, n, n'}_{k_{as}}  \conc \fat{a, c, n, n'}_{k_{cs}} & \multicolumn{3}{c}{\xleftarrow{\hspace*{1.7in}}}
	       & + \fat{a, c, n, n'}_{k_{as}}  \conc  \fat{a, c, n, n'}_{k_{cs}}  \\
	\Downarrow & & 			& & \Downarrow \\
	+ \fat{c \conc n \conc n' \conc k_{ac}}_{k_{as}} \conc \fat{a \conc n \conc n' \conc k_{ac}}_{k_{cs}} & \multicolumn{3}{c}{\xrightarrow{\hspace*{1.7in}}} 
	    & - \fat{c \conc n \conc n' \conc k_{ac}}_{k_{as}} \conc \fat{a \conc n \conc n' \conc k_{ac}}_{k_{cs}}  \\
		& & 		\Downarrow & & \Downarrow \\
		& & 		-\fat{c \conc n \conc n' \conc k_{ac}}_{k_{as}} \conc \fat{n, n'}_{k_{ac}} & \xleftarrow{\hspace*{0.1in}} & + \fat{c \conc n \conc n' \conc k_{ac}}_{k_{as}} \conc \fat{n, n'}_{k_{ac}} \\
	& & 			\Downarrow & & \Downarrow \\
	& & 			+ \fat{n'}_{k_{ac}} & \xrightarrow{\hspace*{0.1in}} & -\fat{n'}_{k_{ac}} \\
	& & 			& & \Downarrow & & \Downarrow \\
	& & 			& & + \fat{c \conc a \conc n \conc n'}_{k_{cs}} & \xrightarrow{\hspace*{0.1in}} & -\fat{c \conc a \conc n \conc n'}_{k_{cs}} \\
	& & 			& & \Downarrow & & \Downarrow \\
	& & 			& & - \fat{c \conc a \conc n \conc n'}_{k_{cs}} \conc \fat{c, a, n, n'}_{k_{as}}
				     & \xleftarrow{\hspace*{0.1in}} & + \fat{c \conc a \conc n \conc n'}_{k_{cs}} \conc \fat{c, a, n, n'}_{k_{as}} \\
	& & 					& & \Downarrow & & \Downarrow \\
	& & 		  & & + \fat{c \conc n \conc n' \conc k_{ac}}_{k_{cs}} \conc \fat{c \conc n \conc n' \conc k_{ac}}_{k_{as}} 
				     & \xrightarrow{\hspace*{0.1in}} & - \fat{c \conc n \conc n' \conc k_{ac}}_{k_{cs}} \conc \fat{c \conc n \conc n' \conc k_{ac}}_{k_{as}} \\
	& & 					& & \Downarrow & & \Downarrow \\
	& & 			& & - \fat{c \conc n \conc n' \conc k_{ac}}_{k_{cs}} \conc \fat{n \conc n'}_{k_{ac}} 
				     & \xleftarrow{\hspace*{0.1in}} & +  \fat{c \conc n \conc n' \conc k_{ac}}_{k_{cs}} \conc \fat{n \conc n'}_{k_{ac}} \\
	& & 					& & \Downarrow & & \Downarrow \\
	& & 			& & + \fat{n'}_{k_{ac}} & \xrightarrow{\hspace*{0.1in}} & - \fat{n'}_{k_{ac}} \\[2ex]
   Server & & Alice (Init) && \Spy && Alice (Resp) 
      \end{array}
    \end{eqnarray*}
    \hrule
    \caption{The attack bundle for the Woo Lam authentication protocol}
    \label{Fig:WLA_penbundle}
  \end{figure}
\end{landscape}

\begin {example}		
  In the attack bundle to the Woo Lam authentication protocol (see
  Fig.~\ref{Fig:WLA_penbundle}), there are several nodes in which
  honest principals receive encrypted messages.  There is no confusion
  in $\Tuple{\SI{r_{\mathit{serv}}}{2}{\beta}}{1}$, because (i)
  $\fat{a \conc n \conc n' \conc k_{ac}}_{k_{cs}}$ originates on a
  penetrator strand, and (ii) $\fat{c \conc n \conc n' \conc
    k_{ac}}_{k_{as}}$ originates in
  $\Tuple{\SI{r_{\mathit{Init}}}{5}{\beta}}{3}$ in the same protocol
  section, and analogously to the canonical bundle.  Similar arguments
  hold for $\Tuple{\SI{r_{\mathit{Init}}}{5}{\beta}}{4}$ and
  $\Tuple{\SI{r_{\mathit{Resp}}}{7}{\beta'}}{4}$.
		
  A cross-protocol confusion occurs in
  $\Tuple{\SI{r_{\mathit{Resp}}}{7}{\beta'}}{5}$, because the
  encrypted message $\fat{c \conc n \conc n' \conc k_{ac}}_{k_{as}}$
  originates in $\Tuple{\SI{r_{\mathit{serv}}}{2}{\beta}}{2}$, which
  is situated in a different protocol section.  It is also a message
  confusion. Although the encrypted message $\fat{c \conc n \conc n'
    \conc k_{ac}}_{k_{as}}$ originates in the attack bundle at
  $\Tuple{\SI{r_{\mathit{serv}}}{2}{\beta}}{2}$, which corresponds to
  $\Tuple{r_{\mathit{serv}}}{2}$ in the canonical bundle, the
  positions in which they occur differ.\hfill\nqed
\end{example}

\subsection{Limits of Changing Messages to Protocol Fix}

We now deal with the situation in which an attacker has already been
able to trick an honest principal into accepting a message $m'$,
instead of an expected message, $m$, for a given protocol.  Therefore,
we want to change the protocol in such a way that the attacker is no
longer able either to construct $m'$ from a collection of eavesdropped
messages, or to pass $m'$ as a camouflaged message for $m$. To
formalize this idea, we first introduce the following definitions.
\begin{definition}[Agents Knowledge]
  Let $s$ be a strand of a bundle $\mathcal{B}$. Further, let
  $\Atoms_{\mathit{Init}}$ and $\keys_{\mathit{Init}}$ respectively be
  the set of atomic messages in $s$ and the set of keys, initially
  known to the agent of $s$. In particular, $\Atoms_{\mathit{Init}}$
  includes all atomic messages that are originating in $s$.
	
  Given a node $\Tuple{s}{i}$ of $s$ then
  $\mathit{pred}(\Tuple{s}{i})$ is the closest negative predecessor
  node of $\Tuple{s}{i}$, if any, and it is undefined, otherwise;
  i.e.~if $\forall 1 \le j < i.\; \sign(\Tuple{s}{j} = +$.
	
  The set of \emph{known keys} $\keys_{\Tuple{s}{i}}$ in a negative
  node $\Tuple{s}{i}$, with $i \le \depth(s)$, is defined by:
  \begin{gather}
    \keys_{\Tuple{s}{i}} =
    \begin{cases}
      \keys_{\mathit{Init}} & \text{ if } \mathit{pred}(\Tuple{s}{i}) \text{ undefined,} \\
      \keys_{\mathit{pred}(\Tuple{s}{i})}\cup(\analz_{\keys_{\mathit{pred}(\Tuple{s}{i})}}(\Tuple{s}{i}))
      & \text{ otherwise.}
    \end{cases}
  \end{gather}
  and the set of \emph{known atomic messages} $\Atoms_{\Tuple{s}{i}}$
  in a negative node $\Tuple{s}{i}$, with $i \le \depth(s)$, by:
  \begin{gather}
    \Atoms_{\Tuple{s}{i}} = 
    \begin{cases}
      \Atoms_{\mathit{Init}} & \text{if }\mathit{pred}(\Tuple{s}{i})\text{ undefined,} \\
      \Atoms_{\mathit{pred}(\Tuple{s}{i})} \cup \Atoms(\analz_{\keys_{\mathit{pred}(\Tuple{s}{i})}}(\{\msg(\Tuple{s}{i})\})) & \text{ otherwise.} 
    \end{cases}
  \end{gather}
\end{definition}

Now, if a protocol has to be changed in order to avoid confusions,
there are two opposing requirements. On the one hand, we would like to
add additional information into individual messages to allow the
recipient to distinguish the original from the faked message. But, on
the other hand, we would not like to change the intended semantics of
the protocol. Our approach is not based on any explicit semantics of
the protocols under consideration, but the semantics is only
implicitly given by the syntactical design of the protocol. When
changing steps of the protocol syntactically, we ought to make sure
that the implicit semantics of the protocol is not changed either. We
try to achieve this goal by changing individual protocol steps only
using conservative extensions and reshuffles of messages.

Suppose, for example, that to patch a protocol, we need to modify an
encrypted term, $t$.  Clearly, upon the reception of $t$, a principal
can decompose it into the subterms that are visible under some known
keys, $\keys$. When patching a protocol by modifying $t$ into $t'$, we
further require that, under $\keys$, $t'$ exposes at least the same
information to the principal as $t$.

\begin{definition}[Message Substitution]\label{def:iesubst}
  A \emph{message substitution} $\sigma$ is a finite list $\langle
  \fat{t_1}_{k_1} \gets \fat{t_1'}_{k_1}, \ldots, \fat{t_n}_{k_n}
  \gets \fat{t_n'}_{k_n} \rangle$ of pairs of encrypted messages, such
  that $\fat{t_i}_{k_i} = \fat{t_j}_{k_j} \implies i = j$.  As
  expected, we define $Dom(\sigma) = \{\fat{t_1}_{k_1}, \ldots,
  \fat{t_n}_{k_n}\}$. 
  Given $t \in \Terms$, $\sigma(t)$ is defined by:
  \begin{gather}
    \sigma(t) = 
    \begin{cases}
      t                              & \text{ if } t \in \Atoms \\
      \sigma(t_1) \conc \sigma(t_2)  & \text{ if } t = t_1 \conc t_2 \\
      \fat{\sigma(t_i')}_{k_i}       & \text{ if } t = \fat{t_i}_{k_i},\,\fat{t_i}_{k_i}\in Dom(\sigma) \\
      \fat{\sigma(t')}_k              & \text{ if } t = \fat{t'}_k,\,\fat{t'}_k\not\in Dom(\sigma) 
    \end{cases}
  \end{gather}
	We extend message substitition $\sigma$ also to sets $\mess$ of messages by 
	$\sigma(\mess) = \{\sigma(t) | t \in \mess \}$.
\end{definition}
\begin{example} \label{ex:mess_subst}
  Let $\sigma = \{ \fat{A \conc \fat{B}_{K}}_{K} \gets
  \fat{\fat{B}_{K} \conc A \conc C}_{K}, \fat{B}_{K} \gets \fat{B
    \conc D}_{K} \}$ then \\
		$\sigma(\fat{A \conc \fat{B}_{K}}_{K} \conc
  \fat{B}_{K}) = \fat{\fat{B \conc D}_{K} \conc A \conc C}_{K} \conc
  \fat{B \conc D}_{K}$.\hfill\nqed
\end{example}

\begin{definition}\label{def:inf-enhancmnt} 
  A message substitution $\sigma$ is \emph{information enhancing}
  wrt.~a set $\mess$ of messages iff $\forall \fat{t_i}_{k_i} \in
  Dom(\sigma). \; \exists \mess' \subseteq \mess. \; \Flat(t'_i) =
  \Flat(t_i) \cup \mess'$.
  A message substitution $\sigma$ is \emph{injective} on a set $\mess$
  of messages iff $\forall t, t' \in \mess. \; \sigma(t) = \sigma(t')
  \implies t = t'$.
\end{definition}
\begin{example}
  The message substitution $\sigma$ in Example \ref{ex:mess_subst} is
  information enhancing wrt. $\{C, D\}$.\hfill\nqed
\end{example}
\begin{definition}
  Let $\sigma_1$, $\sigma_2$ be two message substitutions with
  $Dom(\sigma_1) \cap Dom(\sigma_2) = \emptyset$.  Then the
  \emph{co-substitutions} $\overline{\sigma}_1$ and
  $\overline{\sigma}_2$ of $\sigma_1$, $\sigma_2$ are defined by:
  \[ \overline{\sigma}_1 = \{ \sigma_2(t) \gets \sigma_2(t') \mid t \gets t' \in \sigma_1 \} \text{ and }
  \overline{\sigma}_2 = \{ \sigma_1(t) \gets \sigma_1(t') \mid t \gets t' \in \sigma_2\} \]
\end{definition}

\begin{lemma}\label{lem:iem}
  Let $\sigma_1$ and $\sigma_2$ be two information enhancing message
  substitutions wrt. $\mess_1$ and $\mess_2$, respectively, such that
  $Dom(\sigma_1) \cap Dom(\sigma_2) = \emptyset$. Then, the
  co-substitutions $\overline{\sigma}_1$ and $\overline{\sigma}_2$ of
  $\sigma_1$ and $\sigma_2$ are information enhancing wrt.
  $\sigma_2(\mess_1)$ and $\sigma_1(\mess_2)$, respectively.
\end{lemma}

  \noindent
  \emph{Proof}\
  We prove that $\overline{\sigma}_2$ is information enhancing
  wrt.~$\sigma_1(\mess_2)$; the proof for $\overline{\sigma}_1$ is
  analogous.  Let $\sigma_2 = \langle \fat{s_1}_{k_1} \gets
  \fat{s'_1}_{k_1}, \ldots, \fat{s_n}_{k_n} \gets
  \fat{s'_n}_{k_n}\rangle$ then $\overline{\sigma}_2 = \langle
  \sigma_1(\fat{s_1}_{k_1}) \gets \sigma_1(\fat{s'_1}_{k_1}), \ldots,
  \sigma_1(\fat{s_n}_{k_n}) \gets \sigma_1(\fat{s'_n}_{k_n}) \rangle$.
  Since $\sigma_2$ is information enhancing wrt.~$\mess_2$, we know that
  $\Flat(s'_i) = \Flat(s_i) \cup \mess_2$ and therefore also
  $\sigma_1(\Flat(s'_i)) = \sigma_1(\Flat(s_i) \cup \mess_2) =
  \sigma_1(\Flat(s_i)) \cup \sigma_1(\mess_2)$.  Because $\sigma_1$
  commutes over flattening, we obtain: $\Flat(\sigma_1(s'_i)) =
  \Flat(\sigma_1(s_i)) \cup \sigma_1(\mess_2)$.\hfill$\boxempty$

\begin{lemma}\label{lem:com_subst}
  Let $\sigma_1$, $\sigma_2$ be two information enhancing message
  substitutions wrt. $\mess_1$ and $\mess_2$, respectively, 
	such that $Dom(\sigma_1) \cap Dom(\sigma_2) = \emptyset$.  
	Let $\mess \subseteq \Terms$ be a set of messages, such
  that $\sigma_{1}$ and $\sigma_{2}$ are injective on $\mess$. Then
  there are information enhancing message substitutions
  $\overline{\sigma}_1$ and $\overline{\sigma}_2$ 
	wrt.\ $\sigma_2(\mess_1)$ and $\sigma_1(\mess_2)$ such that
  \[ \forall t\in \mess. \; \overline{\sigma}_1(\sigma_2(t)) = \overline{\sigma}_2(\sigma_1(t)) \]
\end{lemma}

  \noindent
  \emph{Proof}\ 
  Let $\overline{\sigma}_1$ and $\overline{\sigma}_2$ be the
  corresponding co-substitutions of $\sigma_1$ and $\sigma_2$.  By
  Lemma \ref{lem:iem} we know that $\overline{\sigma}_1$ and
  $\overline{\sigma}_2$ are information enhancing message
  substitutions wrt.\ $\sigma_2(\mess_1)$ and $\sigma_1(\mess_2)$,
  respectively.
	
  We prove the push-out by structural induction on the term $t$: Since $\sigma_{1}$ and
  $\sigma_{2}$ are injective on $\mess$, we know that $t \not\in
  Dom(\sigma_1)$ implies $\sigma_2(t) \not\in
  Dom(\overline{\sigma_1})$, and $t \not\in Dom(\sigma_2)$ implies
  $\sigma_1(t) \not\in Dom(\overline{\sigma_2})$, for all $t \in
  \mess$.
  \begin{itemize}
  \item Let $t \in \Atoms$ then $\overline{\sigma}_1(\sigma_2(t)) = t
    = \overline{\sigma}_2(\sigma_1(t))$.

  \item Let $t = t_1 \conc t_2$ then $\overline{\sigma}_1(\sigma_2(t_1
    \conc t_2)) = \overline{\sigma}_1(\sigma_2(t_1)) \conc
    \overline{\sigma}_1(\sigma_2(t_2)) =
    \overline{\sigma}_2(\sigma_1(t_1 \conc t_2))$.

  \item Let $t \in Dom(\sigma_1)$. Hence, $t = \fat{t_i}_{k_i}$ and
    $\sigma_1(t) = \fat{t'_i}_{k_i}$.  Furthermore, $t \not\in
    Dom(\sigma_2)$ and therefore $\sigma_1(t) \not \in
    Dom(\overline{\sigma_2})$. Then $\overline{\sigma}_2(\sigma_1(t))
    =
    \overline{\sigma}_2(\fat{\sigma_1(t'_i)}_{k_i}) 
    =
    \fat{\overline{\sigma}_2(\sigma_1(t'_i))}_{k_i} 
    = \fat{\overline{\sigma}_1(\sigma_2(t'_i))}_{k_i}  
    = \overline{\sigma}_1(\fat{\sigma_2(t_i)}_{k_i})   
    = \overline{\sigma}_1(\sigma_2(\fat{t_i}_{k_i}))   
    = \overline{\sigma}_1(\sigma_2(t))$.

  \item Let $t \in Dom(\sigma_2)$. Analogous to the previous case.

  \item Let $t = \fat{t'}_k \not\in Dom(\sigma_1) \cup Dom(\sigma_2)$.
    Then $\overline{\sigma}_2(\sigma_1(t)) =
    \overline{\sigma}_2(\fat{\sigma_1(t')}_k)$ 
    =
    $\fat{\overline{\sigma}_2(\sigma_1(t'))}_k 
    =
    \fat{\overline{\sigma}_1(\sigma_2(t'))}_k 
    =
    \overline{\sigma}_1(\fat{\sigma_2(t')}_k) 
    = \overline{\sigma}_1(\sigma_2(\fat{t'}_{k}))
    $.\hfill$\boxempty$
  \end{itemize}

\begin{definition}[Message Enhancement]
  Let $m, m'$ be two messages and $\mess$ be a set of atomic messages.
  $m'$ is an \emph{enhancement} of $m$ wrt.~$\mess$, written $m'
  \geq_{M} m$ for short, iff there is an information enhancing message
  substitution $\sigma$ wrt.~$\mess$, such that $\sigma(m) = m'$
  holds.
	
  We write $m' \geq_{M,\sigma} m$ to identify the corresponding
  information enhancing message substitution $\sigma$ with $\sigma(m)
  = m'$.

  We say that $m$ is \emph{equivalent} to $m'$, in symbols $m\equiv
  m'$, iff $m\geq_{\emptyset} m'$. Furthermore, $m >_{M, \sigma} m'$
  iff $m \geq_{M, \sigma} m'$ and $m\not\equiv m'$.
\end{definition}
	
\begin{example}
  $\fat{a\conc b\conc m}_{k_{as}}\equiv\fat{m\conc a\conc
    b}_{k_{as}}$.\hfill\nqed
\end{example}

We shall also demand that the changed message be dissimilar from an
entire set of messages, e.g.~from those used in one or more protocols.
\begin{definition}[Collision Freeness]
  Let $m$ be a message and $M\subseteq\Terms$ be a set of messages.
  Then, $m$ is \emph{collision free with respect to $M$} iff for all
  $m'\in M$ it is the case that $m\not\equiv m'$.
\end{definition}

\subsection{Protocol Repair}
\label{sec:repair}

We have seen that we find bugs in protocols by analyzing a bundle
containing a penetrator strand. Often, our analysis suggests a change
in the structure of a message, and, in that case, it identifies the
node originating such message considering a bundle denoting an
intended protocol run.  With this, we compute the changes to be done
in one such a regular bundle, which are then propagated to the
protocol description. The following definitions aim at a meta-theory
to allow for tracing the consequences of changing all the nodes in the
set of strands caused by a change in a particular node.

\begin{definition}[Adaptation]\label{def:adapt}
  Let $\mathcal{B}$ be a bundle, $v_0$ be a positive node in ${\cal
    B}$ and $\sigma$ be a message substitution. The \emph{adaptation}
  $\Adapt(\mathcal{B}, v_0, \sigma)$ of $\mathcal{B}$ wrt. $v_0$ and
  $\sigma$ is a graph $\mathcal{B}'$ that is isomorphic to
  $\mathcal{B}$; we shall use $\zeta$ to denote the corresponding
  isomorphism that maps the nodes of $\mathcal{B}$ to nodes of
  $\mathcal{B}'$.  The message of each node in $\mathcal{B}'$ is
  defined by $\msg(\zeta(v)) = \msg(v)$ if $v_0 \not\preceq_{\cal B}
  v$ and $\msg(\zeta(v)) = \sigma(\msg(v))$ otherwise.
\end{definition}
Suppose $\mathcal{B}$ is an $\mathcal{R}$-bundle, then the adaptation
$\Adapt(\mathcal{B}, v_0, \sigma)$ can be considered as a canonical
$\mathcal{R}'$-bundle where $\mathcal{R}'$ is the set of renamed
strands embedded in $\Adapt(\mathcal{B}, {v_0, \sigma})$ (c.f.~Section
\ref{sec:roles}).
\begin{definition}
  An adaptation $\Adapt(\mathcal{B}, v_0, \sigma)$ of a canonical
  bundle is \emph{safe} iff $Dom(\sigma)$ originates in $v_0$ and
  $\sigma$ is information enhancing.
\end{definition}

%

\begin{lemma}[Confluence of Adaptations]
  Let $\mathcal B$ be a bundle, $v_1, v_2$ nodes of $\mathcal B$ and
	$\sigma_1, \sigma_2$ information enhancing message substitutions such
	that $Dom(\sigma_1) \cap Dom(\sigma_2) = \emptyset$ and $\Adapt(\mathcal{B}, v_i, \sigma_i)$ ($i =1,2$)
	are safe adaptations. Let $\zeta_i$ be their corresponding node isomorphisms then
	\[\Adapt(\Adapt(\mathcal{B}, v_1, \sigma_1), \zeta_1(v_2), \overline{\sigma_2}) =
	   \Adapt(\Adapt(\mathcal{B}, v_2, \sigma_2), \zeta_2(v_1), \overline{\sigma_1}) \]
	holds	where $\overline{\sigma_1},\overline{\sigma_2}$ are the corresponding co-substitutions of 
		$\sigma_1, \sigma_2$.
\end{lemma}

  \noindent
  \emph{Proof}\ 
  Let $\overline{\zeta_1}$ (and $\overline{\zeta_2}$) be the corresponding
  node isomorphisms for $\Adapt(\Adapt(\mathcal{B}, v_1, \sigma_1),
  \zeta_1(v_2), \overline{\sigma_2})$ (and $\Adapt(\Adapt(\mathcal{B},
  v_2, \sigma_2), \zeta_2(v_1), \overline{\sigma_1})$, repectively).
  We show that $\msg(\overline{\zeta_1}(\zeta_2(v))) =
  \msg(\overline{\zeta_2}(\zeta_1(v)))$ holds for each node $v$ of
  $\mathcal{B}$.
  Obviously $v_k \not\preceq_{\cal B} v$ implies $\zeta_i(v_k)
  \not\preceq_{\cal B} \zeta_i(v)$ and also
  $\overline{\zeta_j}(\zeta_i(v_k)) \not\preceq_{\cal B}
  \overline{\zeta_j}(\zeta_i(v))$ for $i,j,k = 1,2$ and $i \not= j$.
	
  Let $v$ an arbitrary node of $\mathcal{B}$. We do a case analysis on
  the position of $v$ relative to $v_1$ and $v_2$:
  \begin{itemize}
  \item Let $v_1 \not\preceq_{\cal B} v$ and $v_2 \not\preceq_{\cal B}
    v$. Then, $\msg(\overline{\zeta_1}(\zeta_2(v))) = \msg(v) = \msg(\overline{\zeta_2}(\zeta_1(v)))$.
  \item Let $v_1 \not\preceq_{\cal B} v$ and $v_2 \preceq_{\cal B} v$.
	  Then, $\msg(\zeta_1(v)) = \msg(v)$ and $\msg(\zeta_2(v)) = \sigma_2(\msg(v))$. Thus
		$\msg(\overline{\zeta_1}(\zeta_2(v))) = \msg(\zeta_2(v)) = \sigma_2(\msg(\zeta_1(v))) = \msg(\overline{\zeta_2}(\zeta_1(v)))$
  \item Let $v_2 \not\preceq_{\cal B} v$ and $v_1 \preceq_{\cal B} v$.
	  Then, $\msg(\zeta_2(v)) = \msg(v)$ and $\msg(\zeta_1(v)) = \sigma_1(\msg(v))$. Thus
		$\msg(\overline{\zeta_1}(\zeta_2(v))) = \sigma_1(\msg(\zeta_2(v))) = \msg(\zeta_1(v)) = \msg(\overline{\zeta_2}(\zeta_1(v)))$
  \item Let $v_2 \preceq_{\cal B} v$ and $v_1 \preceq_{\cal B} v$.
    Then, $\msg(\zeta_1(v)) = \sigma_1(\msg(v))$ and $\msg(\zeta_2(v)) = \sigma_2(\msg(v))$. Thus,
		$\msg(\overline{\zeta_1}(\zeta_2(v))) = \overline{\sigma_1}(\msg(\zeta_2(v))) = \overline{\sigma_1}(\sigma_2(\msg(v))) 
		= \overline{\sigma_2}(\sigma_1(\msg(v))) = \overline{\sigma_2}(\msg(\zeta_1(v))) = \msg(\overline{\zeta_2}(\zeta_1(v)))$
  \end{itemize}

\section{Patch Methods for Protocols}
\label{sec:rules}

In their seminar paper, \cite{AN96} Abadi and Needham proposed general
guidelines for a protocol design, which cope with the problem of
disambiguating messages in a protocol. In particular, the principle 3,
\emph{agent naming}, prescribes that all agent names relevant for a
message should be derivable either from the format of the message or
from its content. This prevents the reuse of a message in the
corresponding step of a different run of the same protocol.  Principle
10, \emph{recognizing messages and encodings}, deals with the second
source of misuse and demands that a principal should be able to
associate which step a message corresponds to. These principles will
result in different ways to fix a flawed protocol.

We shall now discuss various rules to fix a faulty protocol.  As
already illustrated in Section \ref{sec:approach}, we use a standard
protocol analyzer, like AVISPA, to generate an example of an
interleaving of protocol runs in which a honest principal has been
spoofed. This example is translated into the strand space notation
forming an attack bundle $\mathcal{B}$. Additionally, we translate the
intended protocol run into a canonical bundle $\mathcal{B}^c$.
Comparing both bundles provides the necessary information where and
how we have to change the protocol. The first step is to construct a
coverage for $\mathcal{B}$, representing the different views of the
honest principals on the protocol runs of the attack. The ultimate
goal is to remove all confusions occurring in $\mathcal{B}$ by
small changes of individual protocol steps. Therefore, we will analyze
the various sources for confusions and how they can be avoided
by patching the protocol. The type of patch rule used depends on the
type of confusion (c.f.~Definition~\ref{def:confusion}).
 
A rule (to patch a protocol) is a \emph{patch method}, a tuple
comprising four elements: i) a name, ii) an input, iii) preconditions,
and iv) a patch.  The first element is the \emph{name} of the method,
a string, meaningful to the flaw repair performed by the method. The
second element is the \emph{input}: a bundle ${\mathcal B}$ describing
the attack, a canonical bundle $\mathcal{B}^c$ describing the intended
run of the protocol, and a distinguished node in $\mathcal{B}$ causing
a confusion.  The third element is the \emph{preconditions}, a formula
written in a meta-logic that the input objects must satisfy.  \shrimp
uses these preconditions to predict whether the associated patch will
make the protocol no longer susceptible to the attack.  Finally, the
fourth element is the \emph{patch}, a procedure specifying how to mend
the input protocol.


\subsection{Patching Protocols with a Message Confusion Flaw}

Suppose that a given attack bundle $\mathcal{B}$ has a message
confusion in some node $v \in \mathcal{B}$. This means that a honest
principal accepted a message in some step of the protocol that was
faked by the penetrator. Since the confusion is fatal, the penetrator
reused some encrypted message from another message (possibly from a
different protocol step) that he could not construct himself. Hence,
the problem is that the encrypted messages within the reused message
can be confused with the encrypted message that is part of the
expected message. To fix this vulnerability we have to change one of
these messages to break the similarity. The simplest way of doing this
is to rearrange the information stored in the message. For instance,
we may reverse messages of a concatenation occurring in $\msg(v)$. In
general, we search for a message $m$ such that $\msg(v)\equiv m$, but
$\msg(v)\neq m$. In some cases, and in particular if $\msg(v)$ simply
is an encrypted atomic message, there are no messages $m$ satisfying
these conditions, and we have to extend $\msg(v)$ with an additional
information bit, typically a tag, to resolve the confusion.

\begin{definition}[Message-Encoding Rule]\label{def:message-encoding}
  The \emph{message-encoding rule} is the following transformation
  rule:
  \begin{compactdesc}
  \item[\textbf{Input:}] $\;$ $\mathcal{B}^c$, $\mathcal{B}$ with a
    node $v\in\mathcal{B}$ causing a message confusion with a message
    $t = \msg(v) | \pi$
  \item[\textbf{Preconditions:}] $\;$
    \begin{compactenum}[i)]
			\item $\msg(\theta(v)) | \pi$ originates at $v'$ in $\mathcal{B}^c$ at position $\pi'$,
			\item $v$ belongs to a protocol section with renaming $\beta$
    \end{compactenum}
  \item[\textbf{Patch:}] $\;$ \\
	  Let $t'$ be a message and $\mathcal{B}^{'c}$ a graph such that 
		\begin{compactenum}[i)]
			\item $\sigma = \{ \msg(v')|\pi' \gets t' \}$ is an information enhancing message 
		      substitution wrt. $\Atoms_{v'}$,
			\item $t'$ is collision free wrt.\ $\msg(v'')$ for all nodes $v''$ in $\mathcal{B}^c$, and
			\item $\msg(v)$ is not accepted for $\beta(\sigma(\msg(\theta(v))))$ wrt. $\Atoms_v$ and 
			      the known keys $\keys_v$ in $v$.
    \end{compactenum}
		Then, return $\Adapt(\mathcal{B}^c, v', \sigma)$ as a result of the patch.
  \end{compactdesc}
\end{definition}
\begin{example}\label{ex:mess_conf_1}
  Consider again WMF, which violates Abadi and Needham's design
  principle 10. As illustrated in the attack of
  Example~\ref{Ex:WMF_penbundle}, 
  Alice receives the encrypted message $t = \fat{b \conc t_a \conc
    k}_{k_{as}}$ in $v = \Tuple{\SI{r_{\mathit{resp}}}{1}{\beta'}}{1}$
  which originates in $\Tuple{\SI{r_{\mathit{init}}}{1}{\beta}}{1}$.
  In the canonical bundle, however, the corresponding message
  $\mathsf{\fat{a \conc t_{a+d} \conc k}_{k_{bs}}}$ originates in $v'
  = \Tuple{r_{\mathit{Serv}}}{2}$.  which causes a message confusion
  with $t$. Therefore, the preconditions of the patch are applicable
  and we are looking for a message $t'$ such that $\sigma = \{
  \mathsf{\fat{a \conc t_{a+d} \conc k}_{k_{bs}}} \gets m'\}$ is an
  information enhancing message substitution wrt.\
  $\Atoms_{\theta(v')} = \mathsf{\{a, b, t_a, k\}}$.  Now, suppose
  that the implementation of two messages are equal if and only if the
  messages are equal.  We obtain a suitable message $t' =
  \mathsf{\fat{t_{a+d} \conc a \conc k}_{k_{bs}}}$, by permuting the
  messages of $t$. $t'$ is obviously collision free with all nodes of
  the canonical bundle $\mathcal{B}^c$ and $\fat{b \conc t_a \conc
    k}_{k_{as}}$ is not accepted for $\fat{t_a \conc b \conc
    k}_{k_{as}}$ wrt. $\{t_a, b\}$ and $\{k_{as}\}$.
 
 The resulting bundle $\Adapt(\mathcal{B}^c, v', \sigma)$ looks as follows
  (cf.~the original definition, given in Example~\ref{Ex:WMF_bundle}):
	\begin{eqnarray*}  \sansmath
      \begin{array}{ccccc}
        + a, \fat{b \conc t_a  \conc k}_{k_{as}} &   \xrightarrow{\hspace*{0.1in}} &  - a, \fat{b \conc t_a \conc k}_{k_{as}} \\    
        & &    \Downarrow  \\
        & & + \fat{t_{a+d} \conc a \conc k}_{k_{bs}} &   \xrightarrow{\hspace*{0.1in}} & - \fat{t_{a+d} \conc a \conc k}_{k_{bs}} \\[2ex]
        Init && Serv && Resp
      \end{array}
    \end{eqnarray*}\hfill\nqed
\end{example}
\begin{example}\label{ex:mess_conf_2}
  In case of the Woo Lam $\pi_1$ protocol (see
  Example~\ref{ex:tf_WL1}), 
  there is a message confusion in 
	$v = \Tuple{\SI{r_{\mathit{resp}}}{5}{\beta}}{5}$ with 
	$t = \fat{a \conc b \conc n_b}_{k_{bs}}$, $t$ originates in 
	$\Tuple{\SI{r_{\mathit{resp}}}{5}{\beta}}{4}$ while in the canonical bundle the corresponding message
	$\mathsf{\fat{a \conc b \conc n_b}_{k_{bs}}}$ originates in 
	$v' = \Tuple{r_{\mathit{Serv}}}{2}$ at top level.  
	
	Again, we obtain a suitable message
  $t' = \mathsf{b \conc a \conc n_b}$ by permuting the messages of $t$ such that $\sigma =\{t \gets t'\}$
	is an information enhancing message substitution wrt.\ $\Atoms_{\theta(v')}$.
	$t'$ is obviously collision free with all nodes of the canonical bundle $\mathcal{B}^c$
	and $\fat{a \conc b \conc n_b}_{k_{bs}}$ is not accepted for $\fat{b \conc a \conc n_b}_{k_{bs}}$
	wrt. $\emptyset$ and $\{k_{bs}\}$.
	
	Thus, the resulting bundle $\Adapt(\mathcal{B}^c, v', \sigma)$ looks as follows:
	\begin{eqnarray*} \sansmath  
      \begin{array}{ccccc}
        + a & \xrightarrow{\hspace*{0.1in}} & - a \\
        \Downarrow & & \Downarrow \\
        - n_b & \xleftarrow{\hspace*{0.1in}} & + n_b \\
        \Downarrow & & \Downarrow \\
        +\fat{a\conc b\conc n_b}_{k_{as}} & \xrightarrow{\hspace*{0.1in}} & - \fat{a\conc b\conc n_b}_{k_{as}} \\
        & & \Downarrow \\
        & & +\{\!|a\conc b\conc\fat{a\conc b\conc n_b}_{k_{as}}|\!\}_{k_{bs}} &  \xrightarrow{\hspace*{0.1in}} 
        & - \{\!|a\conc b\conc\fat{a\conc b\conc n_b}_{k_{as}}|\!\}_{k_{bs}} \\
        & & \Downarrow & & \Downarrow \\
        & & - \fat{b \conc a \conc n_b}_{k_{bs}} & \xleftarrow{\hspace*{0.1in}} & +\fat{b\conc a\conc n_b}_{k_{bs}} \\[2ex]
        Init && Resp && Serv  
      \end{array} 
    \end{eqnarray*}\hfill\nqed
\end{example}

Notice that the intruder can elaborate a type flaw attack on a
protocol whenever he is able to make a protocol principal accept a
message of one type, $m$, as a message of another, $m'$.  In that
case, $m$ and $m'$ share the same implementation, $m\approx m'$. Thus,
to remove the protocol type flaw, it suffices to break $m\approx m'$.
So \mencoding~will attempt this by rearranging $m$'s structure while
keeping its meaning intact.  If this operation does not suffice to
break the confusion between $m$ and $m'$, it will then insert into $m$
vacuous terms, tags actually, as in~\cite{tagging}.  Finally, when
incurring on these changes, our methods ensure that the new protocol
message does not clash with some other.  

\subsection{Patching Protocols with a Cross-Protocol Confusion Flaw}

Protocol guarantees are usually implemented via cyphertexts (c.f.~the
rationale behind an authentication test~\cite{auth-tests}). To realize
authentication, the name of the agents that are relevant for the
intended consumption of a cyphertext, namely the originator and the
intended recipients, should be all derivable from the cyphertext
itself (c.f.~principle 3 for protocol design of Abadi and
Needham~\cite{AN96}). When this is not the case, the associated
message, and the protocol itself, has agent naming problems.

When an agent naming problem occurs, a penetrator can reuse the
message corresponding to the $i$-th step of the protocol in some run,
to camouflage that of the $i$-th step but of another run.  Thus, the
penetrator reuses one or more associated message cyphertexts so as to
impersonate the corresponding originator or as to redirect them to an
unintended recipient. From a technical point of view, such a message
lacks sufficient information about the particular protocol run it was
generated for.

We introduce a method designed to fix a faulty protocol with a
cross-protocol confusion flaw. We fix this flaw inserting the names of
the correspondents that are not explicitly mentioned in the message
that has been reused to carry out the attack.
\begin{definition}[Agent-Naming Rule]
  The \emph{agent-naming rule} is the following transformation rule:
  \begin{compactdesc}
  \item[\textbf{Input:}] $\;$ 
    $\mathcal{B}^c$, $\mathcal{B}$ with a node $v\in\mathcal{B}$
    causing a cross-protocol but not a message confusion with a message $t$.

  \item[\textbf{Preconditions:}] $\;$
    \begin{compactenum}[i)]
		\item $t$ originates in $v'\in\mathcal{B}$ at position $\pi$,
    \item $v$ belongs to a protocol section with renaming $\beta$, and
    \item $v'$ belongs to a protocol section with renaming $\beta'$, with $\beta\neq\beta'$
    \end{compactenum}

  \item[\textbf{Patch:}] $\;$\\
	  Let $t'$ be a message such that
    \begin{compactenum}[i)]
		 \item $\sigma = \{ \msg(\theta(v'))|\pi \gets t' \}$ is an information enhancing 
		   message substitution wrt. $\{m \in \Atoms_{\theta(v')} \; | \; \beta'(m) \not= \beta(m) \}$,
	   \item $t'$ is collision free wrt.\ $\msg(v'')$ for all nodes $v''$ in $\mathcal{B}^c$,
    \end{compactenum}
		Then, return $\Adapt(\mathcal{B}^c, \theta(v'), \sigma)$ as a result of the patch.
  \end{compactdesc}
\end{definition}
\begin{example}
  Consider the NSPK protocol in Example~\ref{ex:bundle}. 
  There is a cross-protocol confusion (but not a message confusion) in
  $\Tuple{\SI{r_{\mathit{init}}}{3}{\beta}}{2}$ with $t = \fat{n \conc
    n'}_{k_a}$.  The different protocol sections have the following
  renamings: $\beta = \{\mathsf{a} \gets a, \mathsf{b} \gets
  c,\mathsf{n} \gets n, \mathsf{n'} \gets n', \mathsf{k_a} \gets k_a,
  \mathsf{k_b} \gets k_c \}$ and 
 $\beta' = \{\mathsf{a} \gets a,
  \mathsf{b} \gets b,\mathsf{n} \gets n, \mathsf{n'} \gets n',
  \mathsf{k_a} \gets k_a, \mathsf{k_b} \gets k_b \}$.  Thus, $\pi$ is
  empty and $\msg(\theta(v'))|\pi = \mathsf{\fat{n \conc n'}_{k_a}}$.
	
	Obviously, $\beta$ and $\beta'$ differ
        only\footnote{\onehalfspacing Notice that we are not
          interested in adding agent keys to a message for obvious
          reasons.}  in their renaming of the atomic message
        $\mathsf{b}$, which results in a patched message $t' =
        \mathsf{\fat{n \conc n' \conc b}_{k_a}}$ that is to be
        expected in $\Tuple{\SI{r_{\mathit{init}}}{3}{\beta}}{2}$.
        Therefore, we adjust the protocol using $\sigma =
        \{\mathsf{\fat{n \conc n'}_{k_a}} \gets \mathsf{\fat{n \conc
            n' \conc b}_{k_a}}\}$ and obtain the fixed protocol
        (cf.~the original definition, given in
        Example~\ref{Ex:NSPK_bundle}) by $\Adapt(\mathcal{B}^c,
        \theta(v'), \sigma)$:
	\begin{eqnarray*}  
      \sansmath 
      \begin{array}{ccc}
        +\fat{a \conc n}_{k_b} & \xrightarrow{\hspace*{0.5in}} & -\fat{a \conc n}_{k_b}\\
        \Downarrow & & \Downarrow\\
        -\fat{n \conc n' \conc b}_{k_a}  & \xleftarrow{\hspace*{0.5in}} & +\fat{n \conc n' \conc b}_{k_a} \\
        \Downarrow & & \Downarrow\\
        +\fat{n'}_{k_b} & \xrightarrow{\hspace*{0.5in}} & -\fat{n'}_{k_b} \\[2ex]
        Init && Resp
      \end{array} 
    \end{eqnarray*}\hfill\nqed
\end{example}

\subsection{Patching Protocols with a Replay Protection Flaw}
\label{sec:p9}

The agent-naming rule fails to patch a protocol if the set $\{m \in
\Atoms_{v'} \; | \; \beta(m) \not= \beta'(m)\}$ is empty, which means
that there are two identical copies of an honest strand in the
penetrator bundle contributing to different protocol runs. In other
words, the behavior of this principal is deterministic with respect to
the messages received from its environment. This gives rise to a
\emph{multiplicity attack} in which the penetrator simply replays a
communication from a (pre-)recorded protocol run, and causes an agent
to consider that somebody is trying to set up a simultaneous session,
when he is not~\cite{Lowe97b}.

The \sbinding~rule deals with faulty protocols that contain this kind
of flaw, called \emph{replay protection}. Two example faulty protocols
of this kind are WMF (c.f.~Example~\ref{Ex:WMF_bundle}) and DSSK
(c.f.~Example~\ref{Ex:DSSK_bundle}) protocol. None of these protocols
satisfies \emph{strong authentication} of $B$ to $A$, Both prescribe
the responder, $b$, to react upon an \emph{unsolicited
  test}~\cite{auth-tests}.

In both cases, the responder $b$ participates in the protocol only in
a passive way, by receiving some message. Unless the responder stores
details about each protocol run, it cannot distinguish copies of such
a message replayed by an attacker, from genuine messages in
independent protocol runs. Technically speaking, there are no atoms
that originate on the responder strand allowing it to actively
differentiate individual protocol runs.  \shrimp is equipped with a
repair method that introduces a nonce-flow requirement to fix this
flaw~\cite{AN96,PS00,Lowe97b,auth-tests} (c.f.~principle 7 for
protocol design of Abadi and Needham.)  The idea is to use some
handshake or challenge-response approach to involve the responder
actively in the protocol.
\begin{definition}[Session-Binding Rule]
  The \emph{session-binding rule} is the following transformation rule:
  \begin{compactdesc}
  \item[\textbf{Input:}] $\;$ 

    $\mathcal{B}^c$, $\mathcal{B}$ with a node $v\in\mathcal{B}$
    causing a cross-protocol but not a message confusion with a message $\fat{m}_k$.

  \item[\textbf{Preconditions:}] $\;$
    \begin{compactenum}[i)]
    \item $v$ belongs to a protocol section $\cal T$ with renaming
      $\beta$,
    \item $v'$ belongs to a protocol section ${\cal T}'$ with renaming
      $\beta$ and ${\cal T}' \not= {\cal T}$, and
    \item $\fat{m}_k$ originates in $v'\in\mathcal{B}$ at position
      $\pi$.
    \end{compactenum}

  \item[\textbf{Patch:}] $\;$ Introduce a challenge-response between
    the strand $s$ of $\theta(v)$ and the strand $s'$ of $v''$, where
    $v''$ is a minimal element wrt.~$\preceq_{\mathcal{B}^c}$ in
    $\{\overline{v} \;| \; \overline{v} \preceq_{\mathcal{B}^c}
    \theta(v)\}$.

    Return a bundle $\mathcal{B'} =\Tuple{V'}{(\rightarrow' \cup
      \Rightarrow')}$ with:
    \begin{compactenum}[i)]
    \item $V' = V_{\mathcal{B}^c} \cup \{v_1, v_2, v_3, v_4\}$,
    \item $\rightarrow' \; = \; \rightarrow_{\mathcal{B}^c} \cup
      \{(v_1, v3), (v_2, v4)\}$,
    \item $\Rightarrow' \; = \; \Rightarrow_{\mathcal{B}^c} \cup
      \{(\Tuple{s}{\depth{s}}, v_1), (v_1, v2),
      (\Tuple{s'}{\depth{s'}}, v_3), (v_3, v4)\}$,
    \item $\msg(v_1) = \msg(v_3) = \fat{\Agent(v_1) \conc \Agent(v_3)
        \conc n}_{k}$, and
    \item $\msg(v_2) = \msg(v_4) = \fat{f(n) \conc \Agent(v_3) \conc
        \Agent(v_1)}_{k^{-1}}$.
    \end{compactenum}
    where $n$ is a nonce originating in $v_1$, $f$ is an arbitrary
    injective function on nonces, and either $k = k^{-1} \in
    \keys_{\Tuple{s}{\depth{s}}} \cap \keys_{\Tuple{s'}{\depth{s'}}}
    \setminus \keys_P$, or $k$ and $k^{-1}$ are the public and private
    key of $\Agent(v_3)$.
  \end{compactdesc}
\end{definition}

Obviously, this patch solves the problem of having identical copies of
the same honest strand $s$ in a penetrator bundle. Since the newly
introduced nonce originates in $s$, it also uniquely originates in a
faithful realization of the protocol; i.e.~the honest principal will
generate individual (i.e.~different) nonces for each individual
protocol run.

\begin{example}\label{Ex:WMF_repairbundle}
  Consider the patched Wide-Mouth-Frog protocol in Example
  \ref{ex:mess_conf_1}. After eliminating the message confusion there
  is still an attack possible as illustrated in the following
  penetrator bundle. As Lowe \cite{Lowe97b} has described the attack,
  the penetrator impersonates the server and replays the server's
  response making the responder to believe that a second session has
  been established.
\begin{eqnarray*}
      \begin{array}{ccccccc}
        + a, \fat{t_a \conc b \conc k}_{k_{as}} &   \xrightarrow{\hspace*{0.1in}} &  - a, \fat{t_a \conc b \conc k}_{k_{as}} \\    
        & &    \Downarrow  \\
				& & + \fat{a \conc t_{a+d} \conc k}_{k_{bs}} &   \xrightarrow{\hspace*{0.1in}} & - \fat{a \conc t_{a+d} \conc k}_{k_{bs}} \\
				& & & & \Downarrow  \\
        & & && + \fat{a \conc t_{a+d} \conc k}_{k_{bs}} &   \xrightarrow{\hspace*{0.1in}} & - \fat{a \conc t_{a+d} \conc k}_{k_{bs}} \\
				& & && \Downarrow & & \\
				& & && + \fat{a \conc t_{a+d} \conc k}_{k_{bs}} &   \xrightarrow{\hspace*{0.1in}} & - \fat{a \conc t_{a+d} \conc k}_{k_{bs}} \\[2ex]
        Init && Serv && Spy && Resp
     \end{array}
\end{eqnarray*}
Obviously, there is a cross-protocol confusion in the first node of
the second responder strand with the message $\fat{a \conc t_{a+d}
  \conc k}_{k_{bs}}$. The responder strands belong to different
protocol sections; however, both agree in their renamings $\beta$,
$\beta'$ of the canonical bundle. This prevents the application of the
agent-naming rule and enables the application of the session-binding
rule.  $\Tuple{\mathit{Init}}{1}$ is the only
$\preceq_{\mathcal{B}^c}$-minimal element of $\{\overline{v} \;| \;
\overline{v} \preceq_{\mathcal{B}^c} \Tuple{\mathit{Resp}}{1}\}$ and
we choose to use the increment by one as $f$.
$\keys_{\Tuple{\mathit{Init}}{1}} \cap
\keys_{\Tuple{\mathit{Resp}}{1}} \setminus \keys_P = \{ k, k_{as} \}
\cap \{ k, k_{bs} \} = \{k\}$. Hence, we select $k$ as the encryption
key for the challenge response and we obtain the following patched
canonical bundle:
    \begin{eqnarray*}  \sansmath
      \begin{array}{ccccc}
        + a, \fat{t_a \conc b \conc k}_{k_{as}} &   \xrightarrow{\hspace*{0.1in}} &  - a, \fat{t_a \conc b \conc k}_{k_{as}} \\    
        & &    \Downarrow  \\
        \Downarrow & & + \fat{a \conc t_{a+d} \conc k}_{k_{bs}} &   \xrightarrow{\hspace*{0.1in}} & - \fat{a \conc t_{a+d} \conc k}_{k_{bs}} \\
				  & & & &  \Downarrow  \\
				- \fat{a \conc b \conc n_b}_k  &  \multicolumn{3}{c}{\xleftarrow{\hspace*{0.7in}}}  & + \fat{a \conc b \conc n_b}_k \\
				\Downarrow & & & &  \Downarrow  \\
				+ \fat{n_b + 1 \conc b \conc a}_k  &
                                \multicolumn{3}{c}{\xrightarrow{\hspace*{0.7in}}}  & - \fat{n_b + 1  \conc b \conc a}_k \\[2ex]
        Init && Serv && Resp
      \end{array}
    \end{eqnarray*}\hfill\nqed
\end{example}

\begin{example}
  As another example consider the attack on the DSSK protocol in
  Example \ref{Ex:DSSK_penbundle}.  Again we have two identical copies
  of the responder strand. Similar to the previous example, there is
  cross-protocol confusion in the first node of the second responder
  strand with the message $\fat{b \conc k_{ab} \conc a \conc
    t_s}_{k_{bs}}$. Again, $\Tuple{\mathit{Init}}{1}$ is the only
  $\preceq_{\mathcal{B}^c}$-minimal element of $\{\overline{v} \;| \;
  \overline{v} \preceq_{\mathcal{B}^c} \Tuple{\mathit{Resp}}{1}\}$ and
  we choose to use the increment by one as $f$. With
  $\keys_{\Tuple{\mathit{Init}}{3}} \cap
  \keys_{\Tuple{\mathit{Resp}}{1}} \setminus \keys_P = \{ k_{ab},
  k_{as} \} \cap \{ k_{ab}, k_{bs} \} = \{k_{ab}\}$ we choose $k_{ab}$
  as the encryption key and obtain the following patched DSSK
  protocol:
  \begin{eqnarray*}  \sansmath \scriptsize
    \begin{array}{ccccc}
      & & + a \conc b  & \xrightarrow{\hspace*{0.1in}} & - a \conc b  \\
      & & \Downarrow   & & \Downarrow \\
      & & - \fat{b \conc k_{ab} \conc t_s \conc \fat{b \conc k_{ab} \conc a \conc t_s}_{k_{bs}}}_{k_{as}} & \xleftarrow{\hspace*{0.1in}} &
      + \fat{b \conc k_{ab} \conc t_s \conc \fat{b \conc k_{ab} \conc a \conc t_s}_{k_{bs}}}_{k_{as}}  \\
      & & \Downarrow  \\
      - \fat{b \conc k_{ab} \conc a \conc t_s}_{k_{bs}} & \xleftarrow{\hspace*{0.1in}}  & + \fat{b \conc k_{ab} \conc a \conc t_s}_{k_{bs}} \\
			\Downarrow  & & \Downarrow \\
			+ \fat{a \conc b \conc n_b}_{k_{ab}}  &  \xrightarrow{\hspace*{0.1in}}  & - \fat{a \conc b \conc n_b}_{k_{ab}} \\
	  	\Downarrow & &  \Downarrow  \\
			- \fat{n_b + 1  \conc b \conc a}_{k_{ab}}  &  \xleftarrow{\hspace*{0.1in}}  & + \fat{n_b + 1  \conc b \conc a}_{k_{ab}} \\[2ex]
      Resp && Init && Serv   	
    \end{array} 
  \end{eqnarray*}\vspace*{-0.2in}\hfill\nqed
\end{example}

\subsection{Combining Rules}

As already mentioned in Section \ref{sec:repair}, our approach runs a
protocol verifier and uses its output, in particular the faulty
protocol run, to compute the attack bundle and to select an
appropriate patch rule to fix the protocol.  We apply one rule at a
time and resubmit the patched protocol to the protocol verifier. This
process iterates until either no more flaws are detected by the
verifier, or there are no more applicable rules. Since this approach
constitutes some sort of rewrite system on security protocols, it is
tempting to analyze formal properties of rewrite systems, like
confluence and termination in this setting.

Suppose there is a faulty security protocol containing various flaws.
In the first place, the order in which \shrimp examines the flaws
depends on the order in which attacks on the security protocol are
discovered by the protocol verifier. Given a particular attack bundle,
we may have different messages that allow the penetrator to mount the
attack. Concerning a particular message, at most one of the rules
message-confusion, agent-naming and session binding would be
applicable, because the preconditions of the three rules exclude one
another.  As far as an attack makes use of a faulty implementation and
the attack bundle involves I-traces, there is a choice point between
patching the implementation, such that the type confusion is no longer
applicable or patching the protocol itself so that any type confusion
does not cause any harm.  Both patches are independent of each other,
since they operate on different levels.  If the attacker bundle
reveals several flaws, i.e.~there are various messages that can be
misused to mount an attack, then, lemma~\ref{lem:com_subst} guarantees
that rules based on adaptations like the message-confusion and the
agent-naming rules commute.

There is no strong argument concerning the termination of \shrimp,
although in all our examples (see Section \ref{sec:results}) it was
never the case that our approach did not terminate. One reason is that
rules working on ambiguous messages will disambiguate them with
respect to the entire bundle. Therefore, the number of potentially
ambiguous messages in a protocol will decrease with each application
of such a rule.

\section{Results}
\label{sec:results}

\begin{table}[htb]
  \centering
  \begin{tabular}{lcc}
    \hline
    \textbf{Protocol} & \textbf{Source} & \textbf{Attack Source}\\
    \hline
    Woo-Lam $\pi_1$ & \cite{woolam} & \cite{tagging}\\
    Otway-Rees & \cite{BAN89} & \cite{BAN89}\\
    Neuman-Stubblebine & \cite{NeumanS93} & \cite{Carlsen94}\\
    KP & \cite{roles-crypt-prot} & \cite{roles-crypt-prot}\\
    NSLPK & \cite{Lowe97b} & \cite{tagging}\\
    BAN Yahalom & \cite{BAN89} & \cite{BAN89}\\
    GDOI & \cite{Meadows03} & \cite{Meadows03}\\
    \hline
  \end{tabular}
  \caption{Experimental results: protocols repaired by hand, part II}
  \label{table:validation2}
\end{table}

\begin{table}[htbp]
\begin{center}
\begin{tabular}{lcccccc@{\hspace{-1mm}}cc@{\hspace{-1mm}}ccccc}
\multicolumn{1}{c}{}&\multicolumn{6}{c}{\em before}&&\multicolumn{6}{c}{\em after}\\\cline{1-7}\cline{8-14}
\multicolumn{1}{c}{{\bf Protocol}}& $s$ & $wa_i$ & $sa_i$ & $wa_r$ & $sa_r$ &&$M$&& $s$ & $wa_i$ & $sa_i$ & $wa_r$ & $sa_r$\\ \hline\hline
ASRPC   &\True &\True &\True &\True &\False&  &\sbinA& \ok &\True &\True &\True &\True &\True\\\hline
BAN ASRPC           &\True &\False&\DontCare &\DontCare &\DontCare &  & \an  & \ok &\True &\True &\True &\True &\True\\\hline
CCITTX.509(1)       &\True &\False&\DontCare &\DontCare &\DontCare &  & \an  & \ok &\True &\True &\DontCare &\DontCare &\DontCare       \\
                    &\True &\True &\False&\DontCare &\DontCare &  &\sbinA& \ok &\True &\True &\True &\True &\True \\\hline
CCITTX.509(3)       &\True &\False&\DontCare &\True &\True &  & \an  & \ok &\True &\True &\True &\True &\True \\\hline
DSSK    &\True &\True &\False&\DontCare &\DontCare &  &\sbinA& \ok &\True &\True &\True &\True &\DontCare  \\
                    &\True &\True &\True &\True &\False&  &\sbinB& \ok &\True &\True &\True &\True &\True  \\\hline
NSSK&\True &\True &\DontCare &\True &\False&  &\sbinA& \ok &\True &\True &\DontCare &\True &\True         \\ 
	\hline
DSPK    &\True &\False&\DontCare &\DontCare &\DontCare &  &  \an & \ok &\True &\True &\DontCare &\DontCare &\DontCare         \\ 
                    &\True &\True &\False&\DontCare &\DontCare &  &\sbinB& \ok &\True &\True &\True &\DontCare &\DontCare    \\ \hline 
Kao Chow A.~v1   &\True &\True &\False$\dagger$&\True&\True&&\sbinB&\ok&\True&\True&\True &\True &\True    \\ \hline
KSL                 &\True &\False&\DontCare &\DontCare &\DontCare &  & \me  & \ok &\True &\True &\True &\True &\True     \\ \hline 
NSPK&\False&\DontCare &\DontCare &\True &\True &  & \an  & \ok &\True &\True &\True &\True &\True   \\ \hline
BAN OR      &\True &\False&\DontCare &\DontCare &\DontCare &  & \an  & \ok &\True &\True &\DontCare &\DontCare &\DontCare         \\ \hline 
Splice/AS           &\True &\DontCare &\DontCare &\False&\DontCare &  & \an  & \ok &\True &\True &\DontCare &\True &\DontCare         \\ 
                    &\True &\True &\False&\True &\DontCare &  &\sbinB& \ok &\True &\True &\True &\True &\True   \\ \hline
CJ Splice  &\True &\False&\DontCare &\True &\True &  &\sbinB& \ok &\True &\True &\True &\True &\True \\ \hline
HC Splice   &\True &\DontCare &\DontCare &\False&\DontCare &  & \an  & \ok &\True &\DontCare &\DontCare &\True &\DontCare       \\ \hline 
WMF   &\True &\False&\DontCare &\DontCare &\DontCare &  & \me  & \ok &\True &\True &\DontCare &\DontCare &\DontCare     \\ 
                    &\True &\True &\False&\DontCare &\DontCare &  &\sbinB& \ok &\True &\True &\True &\True &\True\\ \hline
WMF++ $\star$ &\True &\True &\False&\DontCare &\DontCare &  &\sbinB& \ok &\True &\True &\True &\True &\True\\ \hline
ASRPC prune $\star$     &\True &\False&\DontCare &\DontCare &\DontCare &  & \an  & \ok &\True &\True &\DontCare &\DontCare &\DontCare       \\ 
                    &\True &\True &\DontCare &\False&\DontCare &  & \an  & \ok &\True &\True &\DontCare &\True &\DontCare      \\ 
                    &\True &\True &\DontCare &\True &\False&  &\sbinA& \ok &\True &\True &\DontCare &\True &\True      \\
                    &\True &\True &\False&\True &\True &  &\sbinA& \ok &\True &\True &\True &\True &\True\\\hline
WLM      &\True &\False&\DontCare &\DontCare &\DontCare &  & \me  & \ok &\True &\True &\True &\True &\True\\\hline
BAN Yahalom         &\True &\True &\True &\False&\DontCare &  & \me  & \ok &\True &\True &\True &\True &\True \\\hline 
A.~DH $\star$ &\True &\DontCare &\DontCare &\False&\DontCare &  & \an  & \ok &\True &\DontCare &\DontCare &\True &\DontCare       \\ 
                    &\True &\DontCare &\DontCare &\True &\False&  &\sbinA& \ok &\True &\DontCare &\DontCare &\True &\True    \\ \hline
2steps SK $\star$  &\True &\DontCare &\DontCare &\False&\DontCare &  & \an  & \ok &\True &\DontCare &\DontCare &\True &\DontCare     \\ 
                    &\True &\DontCare &\DontCare &\True &\False&  &\sbinA& \ok &\True &\True &\DontCare &\True &\True    \\ 
                    &\True &\True &\False&\True &\True &  &\sbinA& \ok &\True &\True &\True &\True &\True\\  \hline
PS-APG~\cite{PS00} & \True & \True & \DontCare & \False & \DontCare &
& E & & \True & \True & \DontCare & \True & \DontCare\\
                    \hline
\end{tabular}
\caption{Experimental results: protocols repaired automatically, part
  I. Columns before and after are used to convey the properties that
  the associated protocol satsifies, (T)rue and (F)alse (X) means
  property was not tested.) $s$ stands for secrecy, $wa$ for weak
  authentication, and $sa$ stands for strong authentication. The
  subscripts $i$ and $r$ denote the initiator and the responder,
  respectively.}
\label{table:validation1}
\end{center}
\end{table}

Tables~\ref{table:validation1} and~\ref{table:validation2} summarize
our results. We considered \numExp~experiments, of which
\cjlib~involve protocols borrowed from the Clark-Jacob
library;\footnote{\onehalfspacing The Clark-Jacob library comprehends
  50 protocols, 26 out of which are known to be faulty. So our
  validation test set contains all but 6 of these security protocols.
  The faulty protocols that were left out are not susceptible to a
  replay attack.}  \modbyhand~are variants of some of these protocols
(annotated with $\star$); \ktypeflaw~were borrowed from the literature
(these 5 protocols are all known to be susceptible to a type flaw
attack); and \nummod~are protocols output by \shrimp, a
next-generation of an input protocol. Next-generation protocols are
shown in a separate row within the associated entry. Protocol
verification was carried out using AVISPA.

Table~\ref{table:validation1} portrays information about
protocols that were fixed fully automatically. Each row displays the
result of testing a protocol against a (hierarchical) collection of
properties: secrecy, $s$, weak authentication of the initiator, $wa_i$
(respectively responder, $wa_r$) and strong authentication of the
initiator, $sa_i$ (respectively responder $sa_r$), where $wa_i<sa_i$
(respectively $wa_r<sa_r$.) While secrecy has a definite meaning,
authentication is not; however, Lowe's hierarchy of
authentication~\cite{Lowe97} has become the standard reference in the
literature. Not surprisingly, AVISPA's levels of authentication,
namely: weak authentication, and (strong) authentication, correspond
to Lowe's. We refer the reader to~\cite{Lowe97} for the precise
meaning of these terms.

The table separates the verification results for the original
protocol, \emph{before}, and the mended protocol, \emph{after}, as
output by \shrimp. The field value that exists at the intersection
between a protocol $P$ and a property $\phi$ might be either \True,
meaning $P$ satisfies $\phi$, \False, meaning $P$ does not satisfy
$\phi$, or \DontCare, meaning this property was not tested (because
$P$ was not expected to satisfy it.)  Column $M$ specifies the method
that was applied to modify each faulty protocol: message (E)ncoding,
agent (N)aming or session (B)inding.  In all our experiments, the
application of a patch method yielded a revised protocol able to
satisfy the security property that the original one did not. Whenever
applicable, each mended protocol was then further requested to satisfy
the remaining, stronger properties in the hierarchy, thus explaining
why some entries have several runs.  Note that in the discovery of
some attacks we had to specify the possibility of losing a session key
(annotated with $\dagger$.)

\shrimp is thus able to automatically identify a flaw and a successful
candidate patch in \numExito~faulty protocols. Of these experiments,
it applied \numan~times \anaming, \numme~times \mencoding, and
17~times \sbinding. Notice that \shrimp was able to repair 18
protocolos of the \cjlib that were borrowed from Clark and Jacob. The
other two protocols could not be corrected automatically, since
we are currently extending \shrimp so as to incorporate our extension
of the strand spaces to fully account for the theory presented in this
paper. So, the faulty protocols in Table~\ref{table:validation2} have
been tested by hand only.

Further development work is also concerned with automatically
translating a protocol (respectively an attack) from a strand space
notation to AVISPA HLPSL, and back. While most of these translation
can be automated, we anticipate human intervention will be required
for the formulation of security goals, using AVISPA's special
predicates, namely: \emph{witness}, (\emph{weak})
\emph{authentication}, and \emph{authentication\_on}.

We have made \shrimp try to patch the IKEv2-DS protocol, which is part
of AVISPA's library and an abstraction of IKEv2. We found that, upon
an abstraction of the equational issues inherent to the AVISPA attack,
\shrimp successfully identifies a violation to a good practice for
protocol design: the omission of principal names.  While the revised
protocol is up to satisfy strong authentication on the session key,
this patch may be subject to a criticism because IKEv2 was
deliberately designed so that no principal should mention the name of
its corresponding one.  We then deleted $\anaming$ and re-ran our
experiment; this time \shrimp applied \textsf{session binding}
suggesting a protocol similar to IKEv2-DSx, which also is part of
AVISPA's library and attack-free.

\section{Related Work}
\label{sec:related-work}

We now proceed to compare our method against techniques that are rival
in the sense outlined in Section~\ref{sec:intro}.

\subsection{On Protocol Repair}
\label{sec:related-repair}

R.~Choo~\cite{Choo-repair} has also looked at the problem of automated
protocol repair. His development framework applies the SHVT model
checker~\cite{SVHT} to perform a state-space analysis on a
(indistinguishableness-based) model of the protocol (encoded using
asynchronous product automata) under analysis. If the protocol is
faulty, Choo's framework will first check if the associated attack is
captured in the database of attack classes, and then will apply the
repair.  Unfortunately, the attack classes are not formalized, and it
is not clear whether for each class there is an associated repair
method.  \cite{Choo-repair} introduces only one repair method; the
method indiscriminately inserts the names of all the participants
involved in every cyphertext of the protocol. By way of comparison,
for each attack class we provide a formal characterization, amenable
to mechanization, and an specific repair strategy. What is more, if
applied, our naming repair mechanism will selectively introduce only
the agent names required to rule out the attack.  Also, the class of
replay attacks, fully captured by our methods, comprehends the attack
classes involved in~\cite{Choo-repair}.

\subsection{On Protocol Compiling}
\label{sec:related-refinement}

The compiling approach to protocol development takes a protocol that
is weak, in some security sense, and returns other that is stronger,
and made out of modifications of the input one. Example methods in
this vein include~\cite{KatzY03,CortierWZ07,ArapinisDK08}. They are
all based on the seminal work of Canetti on universal composition,
which we describe first as it will be the starting point of our
comparison.

Canetti's Universal composition theorem~\cite{Canetti01} is used to
deduce the security of a complex system, from a proof of the security
of its constituents. For security protocol development, composition
theorems are very attractive, since they assert that if a protocol is
secure, when considered in isolation, then it will remain secure, even
if simultaneously run an unbounded number of times. In particular,
\cite{Canetti01} has shown that for universal composition of a
protocol to hold, it suffices that each of its runs is independent,
and that run independency can be guaranteed prefixing a
pre-established Session ID (SID) into every plaintext message, which
is to be subject of a cryptoprimitive (encryption or signing) by the
protocol. We call this transformation \emph{SID session tagging}.
Although Canetti provides various ways of forming and pre-establishing
a SID, it is common for a SID to be set the concatenation of the
participants' names, together with the fresh nonces such participants
have all previously generated, and exchanged one another (thus the
name \emph{pre-established SID}).

Canetti's proof hinges on the assumption of state disjointness across
all instances of a protocol, thus, ruling out the possibility of
using, for example, long-term keys.  Accordingly, Canetti and
Rabin~\cite{CanettiR02}, later on, showed that Canetti's results still
holds for an, albeit limited, shared randomness, including long-term
keys. Put differently, SID session tagging is enough to provide state
disjointness. 

Yet, K\"usters and Tuengerthal~\cite{KustersT11} have recently shown
that SID session tagging is quite a strong transformation for
separating runs one another. In particular, they established that
universal composition can be achieved on two provisos: first, that the
protocol is secure in the single-session setting, and, second, that it
satisfies a property they call \emph{implicit disjointness}. To put it
another way, real-world security protocols do not typically use a
pre-established SID; therefore, it is necessary to find a further
ordinary condition, namely: implicit disjointness, whereby state
disjointness can be guaranteed. For a protocol $P$ to satisfy implicit
disjointness, two conditions are required. First, participants are not
compromised. Second, conversations in a session, especially decryption
preceeded by encryption, must match. Session matching is, in turn,
guaranteed by an appropriate partnering function. Together, implicit
disjointness and single-session security, resemble Gutmann's notion of
\emph{skeleton}~\cite{skeletons}.  Implicit disjointness can be met
simply by selectively inserting randomness in cryptographic material.

In the remaining of this section, we survey protocol compiling
methods, stemmed from composition theorems.  We shall assume that,
after $k$ previous rounds, participants $a_1,\ldots,a_k$ have
exchanged fresh nonces, $n_1,\ldots,n_k$, to come out with
$SID=a_1;\ldots;a_k;n_1;\ldots n_k$.

Katz and Yung~\cite{KatzY07} first showed how to turn a Key Exchange
(KE) protocol, known to be secure against a passive adversary, into
one that is secure against an active adversary; i.e., into an
authenticated KE protocol. Their method transforms each protocol step,
$j.\;a_i\rightarrow a_{i+1}:\,j;a_i;m$, into:
\begin{eqnarray*}
  j:\;a_i\rightarrow a_{i+1}.\;j;a_i;m;\sqfat{j;m;SID}_{K^-_{a_i}}
\end{eqnarray*}
Later on, Cortier et al.~\cite{CortierWZ07} suggested a transformation
that guarantees beyond authentication, at the cost of assuming a
weaker input protocol that is only functional. Each protocol step,
$j.\;a_i\rightarrow a_{i+1}:\,m$, they change into:
\begin{eqnarray*}
  j:\;a_i\rightarrow a_{i+1}.\;\{|\!m;\sqfat{j;m;SID}_{K^-_{a_i}}|\!\}_{K_{a_{i+1}}}
\end{eqnarray*}
which makes heavy use of cryptoprimitives. Notice that, in particular,
\cite{KatzY07,CortierWZ07}'s methods depend on the creation of new
signing/encryption key pairs, apart from those that already appear in
the original protocol. Finally, Arapinis et al.~\cite{ArapinisDK08}
suggested a straightforward application of Canetti's universal
composition theorem, using SID session tagging, although only
cyphertexts are modified.  Notice that the plus of this method is that
it does not require the creation of new key pairs.

Thus, following the result of~\cite{KustersT11}, all these protocol
refinement techniques have several weaknesses.  Firstly, most assume
the existence and availability of a public key infrastructure.
Secondly, they increase both the round complexity, and the message
complexity: the extra message exchange implements actually a form of
contributary key-agreement protocol; yet, minimizing the number of
communications usually is a design goal.  Thirdly, they involve
additional, indiscriminate computational effort, associated to the
extra encrypted material, thus increasing the communication
complexity.  Fourthly, SID session tagging is as typing tagging (see
review on~\cite{tagging} below), albeit dynamic, and thus poses the
same additional vulnerabilities (the adversary knows SID, and,
therefore, can use it to drive the breaking of long-term keys). And
fifthly, in all these mechanisms, it is assumed that the participants
are all honest: this is restrictive, since it is necessary to consider
attacks where the penetrator is part of the group, or even where
several participants collude to exclude one or more participants, as
in~\cite{MKR05}. These properties are all in contrast with our
approach, since: we do not require extra key pairs; we selectively
include extra rounds, by means of key confirmation; we selectively
insert extra bits of information on an specific message; and we
minimize the use of tags.

Concluding, even the less heavily cryptographic dependant
mechanism of protocol refinement~\cite{ArapinisDK08} would `repair',
for example, NSPK (see Example~\ref{Ex:NSPK_bundle}) as follows:
\begin{eqnarray*}
  a \rightarrow b & : & n\\
  b \rightarrow a & : & n'\\
  a \rightarrow b & : & \fat{a;b;n;n';a;n_a}_{K_b}\\
  b \rightarrow a & : & \fat{a;b;n;n';n_a;n_b}_{K_a}\\
  a \rightarrow b & : & \fat{a;b;n;n';n_b}_{K_b}\\
\end{eqnarray*}
This increases the number of rounds, and triples the size of the
messages to be ciphered. Also, notice that Choo's approach would not
include the extra first two steps, but would tag each message to be
cyphered using $SID=a;b$. By way of comparison, our repair mechanism
would yield NSLPK (see Example~\ref{ex:nslpk}.)

\subsection{On Further Strategies for Protocol Repair, Based on Old Principles}

When considering an automatic protocol repair mechanism like ours, one
wonders whether there is an upper bound as to the information that
every message should include to avoid a replay. If there is one, we
could simply ensure that every message conforms it previous to any
verification attempt. Carlsen~\cite{Carlsen94} has looked into this
upper bound. He suggested that to avoid replays every message should
include five pieces of information: protocol-id, session-id, step-id,
message subcomponent-id and primitive type of data items. In a similar
vein, Malladi et al.~\cite{MAH02} suggested one should add a
session-id contributed to by every participant to any cyphertext in
the protocol, as in~\cite{Canetti01}, and Aura~\cite{Aura97} suggested
one should also use several crypto-algorithms in one protocol and hash
any authentication message and any session key. Protocol designers,
however, find including all these elements
cumbersome.\footnote{\onehalfspacing This is indeed supported by, for
  example, \cite{KustersT11}, who state that while the use of a
  pre-established SID is a good design principle, real-world security
  protocol do not adhere to it, at least not in the explicit way
  portrayed by this principle.}  By comparison, \shrimp only inserts
selected pieces of information considering the attack at hand but may
add steps to the protocol if necessary to fulfill a stronger security
property.  What is more, it follows, from Canetti's universal
composition theorem, that these all principles are subsumed by SID
session tagging, which inserts even less information.


\subsection{On Protocol Synthesis}

Complementary to ours is the work of Perrig and Song~\cite{PS00}, who
have developed a system, called APG, for the synthesis of security
protocols. The synthesis process, though automated, is generate and
test: APG generates (extends) a protocol step by step, taking into
account the security requirements, and then discards those protocols
that do not satisfy them.  APG is limited to generate only 3-party
protocols (two principals and one server). As a reduction technique,
it uses an impersonation attack and so rules out protocols that
(trivially) fail to provide authenticity. The main problem to this
tool is the combinatorial explosion (the search space is of the order
$10^{12}$ according to the authors).

\subsection{On Type Flaw Attacks}

In the past, several approaches have been developed to deal with
potential type flaw attacks. This work can be classified into two
different areas: The first area is concerned with changing the
representation of messages to prevent type flaw attacks in the first
place. Heather et al. \cite{tagging} use a tagging of messages to
identify the type of a message in a unique way. Considering original
messages combined with their tagging as new atomic entities, such
messages constitute a generated algebraic datatype with
non-overlapping ranges of the constructors satisfying the most
important properties of the abstract message theory. Since tags
themselves will reveal information about a message to a penetrator
there are several refinements of this tagging approach to minimize the
set of subterms of a message that have to be tagged (e.g.
\cite{type-guessing,MAH02}).

The second area is concerned with the verification of protocols that
may contain type-flaw attacks.  A prominent approach is to replace the
standard representation of messages as a freely generated datatype by
a more involved datatype dealing with equations between constructor
terms. As a consequence, terms representing messages have to be
unified modulo a theory modeling the equality relation on messages.
While this approach assumes that only entire messages can be
interpreted in different ways, Meadows \cite{Meadows02,Meadows03}
investigated the problem that the implementation of a message could be
cut into pieces and one of such pieces might be used to mock another
message, e.g.~some part of a bit-string representing an encrypted
message is reused as a nonce. In her model, messages are represented
as bit-streams. Based on an information flow analysis of the protocol
and the knowledge how abstract messages are represented as bit-streams,
probabilities are computed as to how likely a message could be guessed
(or constructed) by a penetrator. This is in contrast to our
possibilistic approach, in which we abstract from unlikely events,
e.g.~that independently guessing a key and a nonce will result in
messages that share the same implementation.  This is reflected in our
notion of originating messages occurring on \textbf{I}-traces. The
strand space approach excludes protocol runs in which messages are not
uniquely originating resulting in a possibilistic approach in a
somewhat idealized world.  In our approach, for instance, a nonce can
be only camouflaged by a message, which itself is camouflaged at some
point with the help of the same nonce
(c.f.~definition~\ref{def:origin}).

We separate the protocol level operating on abstract messages from its
implementation level. This is in contrast to many other approaches
that encode implementation details in an equality theory on (abstract)
messages.  The benefit is that we can use (arbitrary) algebraic
specifications to formalize properties (in particular equality and
inequality) of the message implementation.  This knowledge about the
implementation is used to verify side conditions of \textbf{I}-traces.
In order to apply \textbf{I}-traces in the penetrator bundle we have
to make sure that both messages of the trace share the same
implementation. This is a task that can be given to an automated
theorem prover or to specialized deduction system incorporating domain
specific knowledge (e.g. SMT-provers).

\section{Conclusions}
\label{sec:conclusions}

We presented a framework for patching security protocols. The
framework is formalized based on the notion of strand spaces, which we
have extended to deal also with type flaw attacks. Given a
specification for the implementation of messages, an additional
\textbf{I}-trace rule allows the penetrator to interchange messages
that share the same implementation. While the original purpose of this
extension was to provide a uniform representation language for
protocol runs (potentially containing type flaw attacks), it is also
interesting to investigate how the verification methodology of strand
spaces can be lifted to our extended approach.

Based on this formalization, we classify the situations in which a
penetrator can reuse messages communicated by honest principals to
mount attacks. This gives rise to various patch rules, which cope with
different types of flaws. It is interesting to see that there are
typically two alternative ways to overcome type flaw attacks. On the
one hand, we can change the implementation in order to avoid the
particular equalities on implemented messages; and, on the other hand,
we can change the protocol on the abstract level, in order avoid
situations in which one can exploit these properties in the
implementation. The framework has been implemented in the \shrimp
system and successfully evaluated in a large set of faulty security
protocols.

\begin{center}
  \small
  \textbf{Acknowledgements}
  \normalsize
\end{center}
This work has been supported by the Deutsche Forschungsgesellschaft
DFG under contract Hu737/4-1 and the Consejo Nacional de Ciencia y
Tecnolog\'{\i}a CONACyT in Mexico under contract 121596.

\bibliography{verify}
\bibliographystyle{plain}

\end{document}